\documentclass[journal]{IEEEtran}
\usepackage{changes}%[final]
\usepackage{amsmath,amsfonts}
\usepackage{algorithmic}
\usepackage{algorithm}
\usepackage[caption=false,font=normalsize,labelfont=sf,textfont=sf]{subfig}
\usepackage{textcomp}
\usepackage{stfloats}
\usepackage{url}
\usepackage{verbatim}
\usepackage{hyperref}
\usepackage{xcolor}
\usepackage{graphicx}
\usepackage{amssymb,amsthm,epsfig,epstopdf,array}
\usepackage{float}
\usepackage{booktabs}
\usepackage{enumitem}
\usepackage{lscape}
\usepackage{siunitx}
\usepackage{pifont}% http://ctan.org/pkg/pifont
\usepackage[switch]{lineno}
\usepackage{multirow}
\usepackage{tikz}
\usetikzlibrary{arrows}
\theoremstyle{plain}

\begin{document}
\title{DCA: Delayed Charging Attack on the Electric Shared Mobility System}

\author{\IEEEauthorblockN{Shuocheng Guo,
Hanlin Chen,
Mizanur Rahman~\IEEEmembership{Member,~IEEE,} and
Xinwu Qian}
\thanks{Shuocheng Guo, Minzanur Rahman, and Xinwu Qian \textit{(Corresponding author)} are with Department of Civil, Construction and Environmental Engineering, The University of Alabama, Tuscaloosa, AL 35487, USA (e-mail: \{sguo18,mizan.rahman,xinwu.qian\}@ua.edu)}
\thanks{Hanlin Chen is with Lyles School of Civil Engineering, Purdue University, West Lafayette, IN 47907, USA (e-mail: chen1368@purdue.edu)}
\thanks{\textit{Accepted by} IEEE Transactions on Intelligent Transportation Systems, 2023.}}

\maketitle
\begin{abstract} 
An efficient operation of the electric shared mobility system (ESMS) relies heavily on seamless interconnections among shared electric vehicles (SEV), electric vehicle supply equipment (EVSE), and the grid. Nevertheless, this interconnectivity also makes the ESMS vulnerable to cyberattacks that may cause short-term breakdowns or long-term degradation of the ESMS.
This study focuses on one such attack with long-lasting effects, the Delayed Charge Attack (DCA), that stealthily delays the charging service by exploiting the physical and communication vulnerabilities.
To begin, we present the ESMS threat model by highlighting the assets, information flow, and access points. We next identify a linked sequence of vulnerabilities as a viable attack vector for launching DCA.
Then, we detail the implementation of DCA, which can effectively bypass the detection in the SEV's battery management system and the cross-verification in the cloud environment.
We test the DCA model against various Anomaly Detection (AD) algorithms by simulating the DCA dynamics in a Susceptible-Infectious-Removed-Susceptible process, where the EVSE can be compromised by the DCA or detected for repair.
Using real-world taxi trip data and EVSE locations in New York City, the DCA model allows us to explore the long-term impacts and validate the system consequences.
The results show that a 10-min delay results in 12-min longer queuing times and 8\% more unfulfilled requests, leading to a 10.7\% (\$311.7) weekly revenue loss per driver. With the AD algorithms, the weekly revenue loss remains at least 3.8\% (\$111.8) with increased repair costs of \$36,000, suggesting the DCA's robustness against the AD.
\end{abstract}


\section*{Nomenclature}
\addcontentsline{toc}{section}{Nomenclature}
\begin{IEEEdescription}[\IEEEusemathlabelsep\IEEEsetlabelwidth{$V_1,V_2,V_3$}]
\item[AD] Anomaly Detection
\item[BMS] Battery Management System
\item[CAN] Control Area Network
\item[CC-CV] Constant-Current-Constant-Voltage
\item[CCS] Combined Charging System
\item[CSMS] Charging Station Management System
\item[Ctrl\&Comm.] Control\&Communication
\item[DCA] Delayed Charging Attack
\item[DCFC] DC Fast Charging
\item[DDoS] Distributed Denial of Service
\item[DSO] Distribution System Operators
\item[ECU] Electronic Control Units
\item[EM] Expectation-Maximization
\item[EMS] Energy Management System
\item[ESMS] Electric Shared Mobility System
\item[EVSE] Electric Vehicle Supply Equipment
\item[FDIA] False Data Injection Attack
\item[GMM] Gaussian Mixture Model
\item[HMI] Human Machine Interface
\item[IF] Isolation Forest
\item[KLD] Kullback-Leibler Divergence
\item[Kmeans] K-Means Clustering
\item[MitM] Man-in-the-Middle
\item[MTTR] Mean-Time-To-Repair
\item[NYC] New York City
\item[OCPI] Open Charge Point Interface
\item[OCPP] Open Charge Point Protocol
\item[PCC] Principal Component Classifier
\item[PLC] Power Line Communication
\item[SEV] Shared Electric Vehicle
\item[SIRS] Susceptible-Infectious-Removed-Susceptible
\item[SoC] State-of-Charge
\end{IEEEdescription}

\section{Introduction}

\begin{figure}[h]
    \centering
    \includegraphics[width=1.0\linewidth]{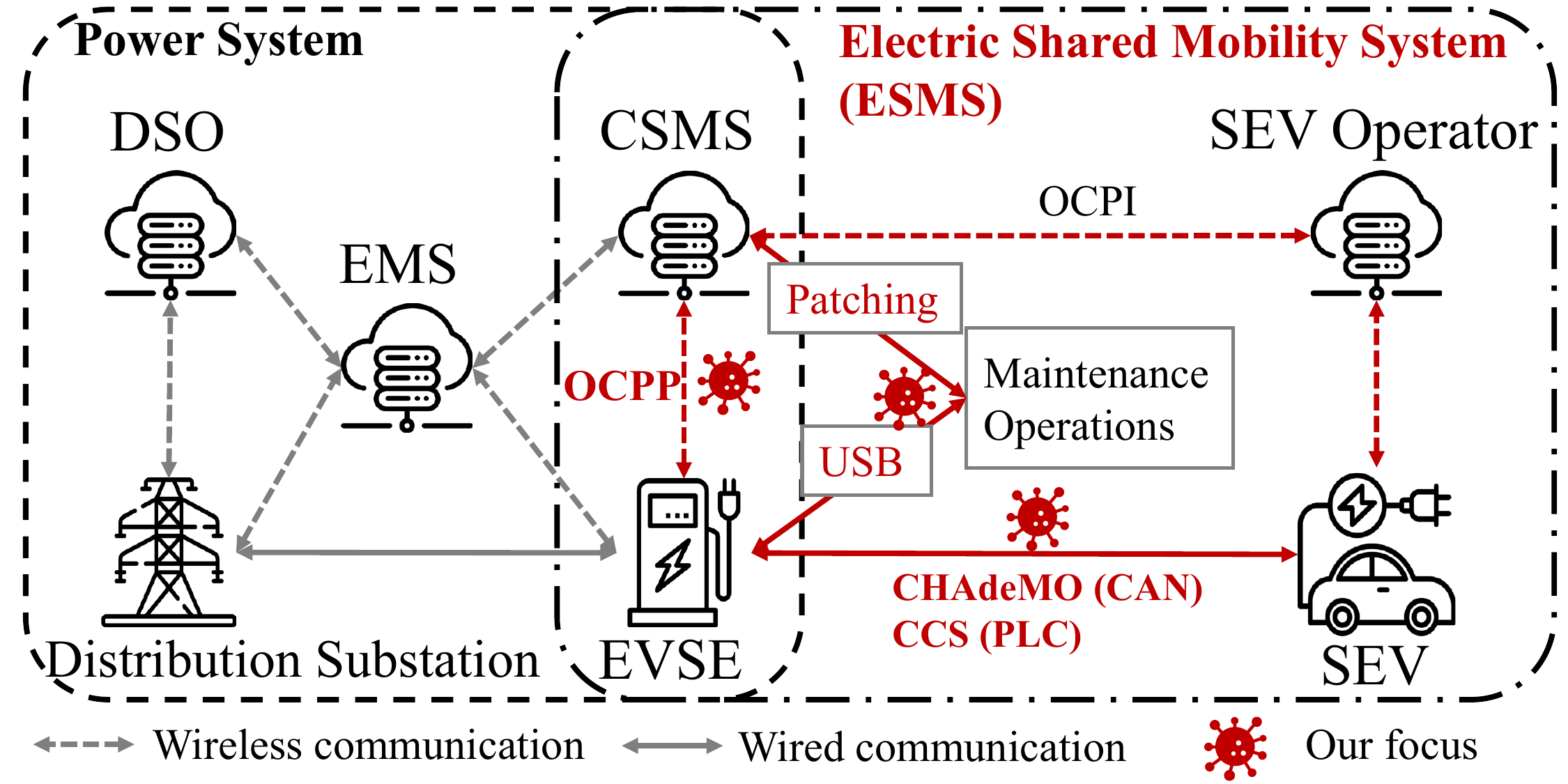}
    \caption{Communication framework in the ESMS (DSO: Distribution System Operators; EMS: Energy Management System)} 
    \label{fig:layout}
\end{figure}

Electrifying the fleet for shared mobility service is a promising direction to lower operation costs and reduce greenhouse gas emissions~\cite{qian2020impact,jenn2020emissions}.
As an example, Shenzhen, China has a fully-electrified taxi fleet by the end of 2019~\cite{lei2022understanding}. Moreover, New York City (NYC) will embrace 100\% electrification of its for-hire vehicles fleet by 2030~\cite{nyc_electric_fleet2030}, which further requires the deployment of over 1,750 DC-Fast charging ports~\cite{moniot2022estimating}.
For large-scale electric shared mobility systems (ESMSs), the essence of the efficient operation is the optimal scheduling of charging and mobility services, which requires seamless interconnections among the major assets in the ESMS (see Fig.~\ref{fig:layout}), including the shared electric vehicles (SEVs), EV supply equipment (EVSE), charging station management system (CSMS), and EVSE and SEV operators, facilitated by multiple communication protocols, e.g., Open Charge Point Protocol (OCPP)~\cite{oca2022oca} and Open Charge Point Interface (OCPI)~\cite{ev2022ocpi}.
However, these communication pathways also present vulnerabilities that can be exploited to compromise the SEVs and EVSE~\cite{kohler2022end,cybersecurity2021multiple,garofalaki2022electric}, 
resulting in the battery charging controller malfunctions and inconsistent changing outcomes for SEVs, thus disrupting the coordination of charging schedules across the entire fleet. This will translate into local congestion at EVSEs, excessive downtime for vehicle supply, and degradation of system performances or even catastrophic failure of the entire mobility system.  
Considering the vulnerabilities and significant consequences above, this study will take the first step to investigate an attack model that can disrupt ESMS and understand its long-term impacts on operational dynamics.

The cybersecurity issues have been extensively studied in the fields of smart grid~\cite{li2017cybersecurity,liu2022false,su2023optimal}, Internet-of-Things~\cite{lu2018internet}, mobility-as-a-service systems~\cite{thai2016resiliency}, traffic signal control
systems~\cite{feng2022cybersecurity}, cyber-physical systems~\cite{duo2022survey,zhao2021decentralized,zhang2021deep}, intelligent transportation systems~\cite{javed2022integration}, and more recently, the ESMS~\cite{mousavian2017risk,acharya2020public,gumrukcu2022impact,johnson2022review,muhammad2023emerging}. 
While the ESMS offers improved efficiency with coordinated charging and dispatching, it also inherits the cyber threats from its assets, including SEVs, EVSE, and the communication interfaces with the CSMS and SEV operator. As such, the ESMS can be subject to new forms of cyberattacks due to the increased system dependencies. 

As depicted in Fig.~\ref{fig:layout}, we showcase a potential attack vector represented by red lines for a distinct type of False Data Injection Attack (FDIA), known as the Delayed Charging Attack (DCA). The DCA aims to stealthily delay the charging process of the SEV fleet while maintaining the safety requirements of the SEV (e.g., avoiding excessive current). This attack can be accomplished by compromising the EVSE via a physical access point (i.e., USB port) or intercepting the communication between the SEV and EVSE by patching via the OCPP during the maintenance operations. The compromised EVSE then injects falsified State-of-Charge (SoC) data into the SEV via the CHAdeMO or CCS physical connections. As a result, the SEV is spoofed to accept a reduced charging rate (e.g. lower current or voltage), which extends the charging time to reach the target SoC. The compromised EVSE and SEV operator bypass cross-verification achieved by the OCPI by uploading constant charging log information to the cloud environment (i.e., CSMS and SEV operator).
For full steps of implementation, see Section~\ref{sec:dca_implementation}.

Unlike other types of attacks (e.g., denial-of-service) which cause an entire breakdown in one shot, the DCA is designed as a stealthy cyberattack with long-term consequences. In this regard, the DCA is likely to be overlooked for two reasons: (1) the DCA will not lead to an immediate collapse of the ESMS, but instead a gradual degradation over time, and (2) the anomalies resulting from the DCA are challenging to detect, particularly in the ESMS, where outliers (e.g., extended charging duration) can be indistinguishable due to the large variation in charging duration (which can range from several minutes to 1-2 hours) across all charging activities.

To the best of our knowledge, no prior studies have investigated the DCA on the large-scale ESMS. This study aims to address this gap by exploring the system-level consequences of the DCA and evaluating the DCA's efficacy and robustness under different length of delayed charging service and various Anomaly Detection (AD) techniques. 
Specifically, we define the robustness of DCA as its ability to maintain the performance even in the presence of AD algorithms, which highlights the DCA's effectiveness and stealthiness. 
This study will begin by presenting the ESMS threat model and identifying a linked sequence of potential vulnerabilities, which forms a viable attack vector for launching DCA. Then, we model the DCA dynamics based on the Susceptible-Infectious-Removed-Susceptible (SIRS) process. Detection strategies are incorporated by comparing widely adopted AD techniques, which can detect the malfunctioning EVSE based on the deviation of the normal system performance. Those malfunctioning EVSE will revert to susceptible state upon repair.

Our major contributions are summarized as follows: 
\begin{itemize}
    \item We present a threat model for ESMS with detailed assets, information flow, and access points. Through our analysis, we identify various vulnerabilities that can be exploited as access points for launching DCA, including physical entry points (e.g., USB ports) and communication interfaces (e.g., signal exchanges between EVSE and SEV, software updates).
    \item We develop a novel DCA model that encompasses attack vectors, potential consequences for the ESMS, and implementation details. We employ different anomaly detection strategies to demonstrate the robustness of the DCA.
    \item We develop a high-fidelity simulation platform that uses real-world data to assess the long-term effects of the DCA on a large-scale ESMS and validate the system consequences of this attack.
\end{itemize}

The rest of the paper is organized as follows. Section~\ref{sec:literature_review} describes ESMS threat model and examines the viability of the DCA model. Section~\ref{sec:model} proposes the SIRS-based DCA model and AD algorithms, and Section~\ref{sec:simulator} introduces the key components in our high-fidelity simulation platform. Section~\ref{sec:experiments} presents the scenario design and parameter assumption, followed by numerical experiments in Section~\ref{sec:results}. Section~\ref{sec:conclusion} concludes our paper.

\section{Threat Model for the ESMS}\label{sec:literature_review}

In this section, we present the threat model of the ESMS by highlighting the major assets, information flows, and access points. We make a significant contribution by streamlining a generic SEV-EVSE threat model~\cite{jay2019securing} to only include the essential components for a minimal implementation of the large-scale ESMS while preserving the same properties as in the original model. By extending the framework to the system level, we demonstrate the communications between a large-scale SEV fleet and the SEV service provider and the potential cascading effects among the SEVs, which is a novel contribution compared to the single-vehicle version threat model. Moreover, we focus on specific cyber-physical threats that occur in the real-world, rather than providing a generic description of possible attack approaches, which allows us to better understand the uniqueness of the ESMS and provides prerequisites for the development of the DCA.

\begin{figure*}[ht]
    \centering
    \includegraphics[width=0.9\linewidth]{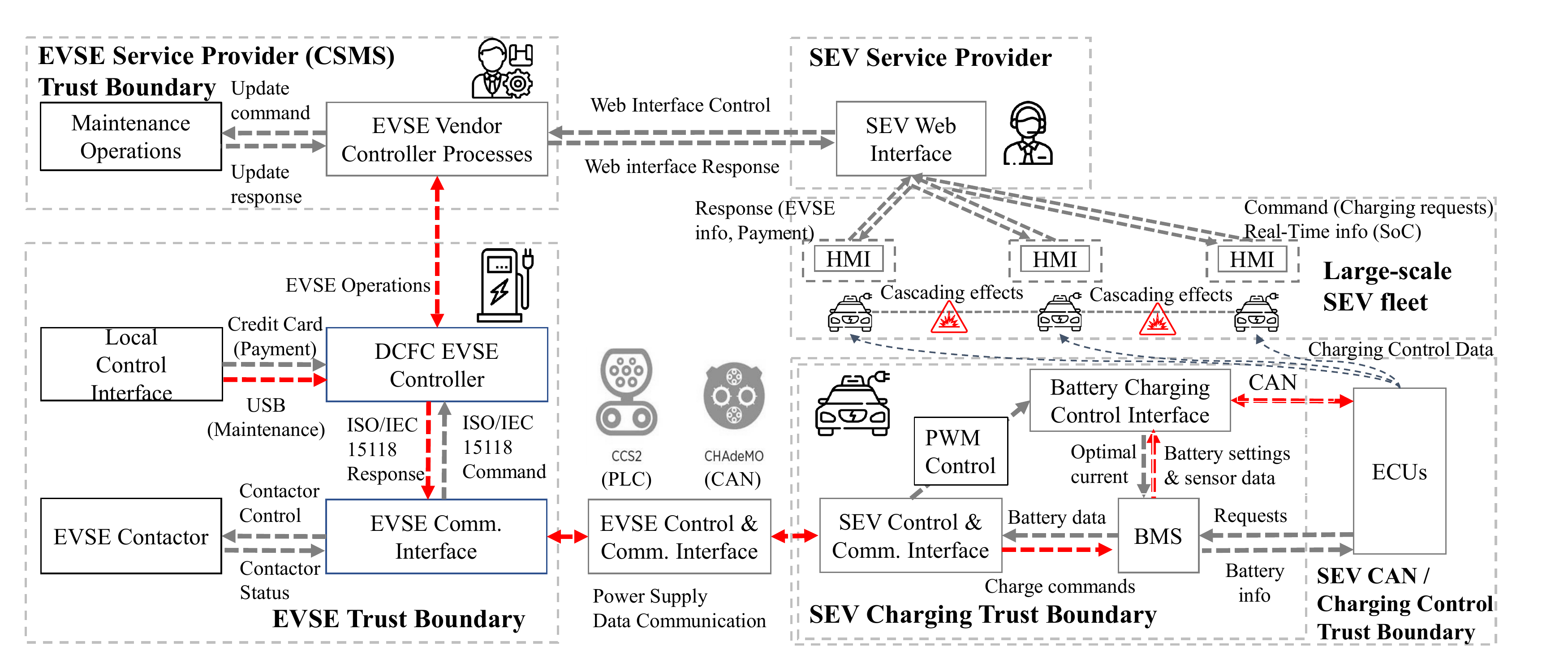}
    \caption{A minimal implementation of ESMS, adapted from the STRIDE threat model proposed by the Sandia National Laboratories~\cite{jay2019securing}. (The red arrows indicate the communication channels that carry falsified data, and the grey arrows represent other communication channels.)}
    \label{fig:system_overview}
\end{figure*}

Our threat model, shown in Fig.~\ref{fig:system_overview}, provides a minimal implementation of the electric shared mobility service operated by a single EVSE and SEV service provider. 
It includes four types of trust boundaries, the SEV fleet, and the SEV service provider. Information flows are detailed between the major assets and interfaces, including EVSE or SEV control \& communication interfaces, connectors, battery management system in the SEV, and the external entities (e.g., SEV and EVSE service providers).

\subsection{Major Assets}
We summarize the major assets in the ESMS threat model as follows:
\begin{itemize}% [leftmargin=*]
    \item \textbf{CSMS}: The CSMS enables SEV drivers and EVSE operators to control and monitor the EVSE remotely, including charging record-keeping, scheduling, and user authentication~\cite{nasr2022power}.
    \item \textbf{EVSE}: The EVSE consists of four major components: DC Fast Charging (DCFC) EVSE controller, EVSE communication interface, EVSE Contactor, and local control interface. The main function of the EVSE is to transport the electric power from the power grid to the SEV battery. 
    \item \textbf{DCFC EVSE controller}: The DCFC EVSE controller processes the exchanged data from the local control interface (e.g., payment information) and EVSE communication interface (e.g., SEV battery data). It processes the real-time information between the EVSE and the EVSE vendor controller in the CSMS, e.g., availability status. In addition, the EVSE controller also supports the maintenance service, which can be conducted by either physical access (e.g., USB) or over-the-air service via OCPP from the CSMS (e.g., patch and software update).
    \item \textbf{EVSE communication interface}: The EVSE communication interface communicates with the EVSE Contactor for power supply, obtains charging requirements from the EVSE control \& communication interface, and sends it to the DCFC EVSE controller~\cite{texas2022chargingstation}. 
    \item  \textbf{EVSE control\&communication interface}: The EVSE control\&communication interface is typically the only physical link between the EVSE and SEV, which is embedded in the charging connector. The most popular types of DC Fast charging connectors in the U.S. include the CHAdeMo, Combined Charging System (CCS), and Tesla charger~\cite{afdc2022chargingspeed}. The connector consists of power lines, control lines, control pilots (i.e., power line communication (PLC) in CCS, or CAN buses in CHAdeMO), enabling power supply, analog control, and data communication, respectively~\cite{jar2016rapid}.
    \item \textbf{Battery Management System (BMS)}: BMS controls the status of the SEV battery within the specified safe operating conditions~\cite{cheng2010battery}. For instance, it monitors the real-time voltage, current, and battery temperature to avoid excessive current or overheating~\cite{jar2016rapid}.
    \item \textbf{SEV control \& communication interface}: SEV battery charging interface collects the battery data information from the BMS and sends the EVSE's configuration to the BMS for a compatibility check~\cite{jar2016rapid}. %\added{Where do you get this information?}
\end{itemize}

\subsection{Information Flows in SEV Charging Process}

The ESMS relies heavily on seamless communication for an efficient coordination of dispatching and charging needs. In particular, a full cycle of charging service, from charging reservations to completion, requires various communication exchanges between the SEV driver, EVSE, and CSMS. For instance, the SEV driver must initiate a charging request with the desired SoC or charging duration, which can be accomplished through the charging app (e.g., EVgo~\cite{evgo2022reserve} and EVmatch~\cite{evmatch2022find}) or human-machine interface (HMI).
Before the charging starts, several rounds of confirmation will proceed via the analog control lines for a compatibility check (e.g., SEV battery and charger parameters)~\cite{jar2016rapid}.
During the charging process, numerous communication exchanges take place between the EVSE and SEV regarding power supply and battery conditions~\cite{jar2016rapid} (e.g., maximum voltage to stop charging, target voltage, battery capacity, and maximum admissible current of the EVSE and SEV). 
Furthermore, the BMS continuously calculates the optimal charging current based on the current SoC, battery condition, and temperature. Upon reaching the target SoC or charging duration, the BMS sends a signal to the EVSE to end the charging process.

\subsection{Access Points}\label{sec:access_points}%%%%Attack surface
The wireless communication and physical entries in the ESMS open a wide attack surface, which can be exploited as access points to disrupt the charging process. We briefly summarize three types of access points in the ESMS as follows (for a comprehensive review, see Johnson et al.~\cite{johnson2022review}).
\begin{itemize}
    \item \textbf{Control Area Network (CAN)}: The initial design purpose of CAN was to ensure communication performance under a complex electromagnetic environment, which does not include consideration for cybersecurity~\cite{currie2017hacking}. The communication mechanism within CAN utilizes a broadcasting mechanism, enabling eavesdropping attacks.
    As seen in Fig.~\ref{fig:system_overview}, 
    the CAN buses cover the major components in the EVSE and SEV through the EVSE control \& communication interface,
    where the transmitted messages can be modified and broadcast to all covered electronic control units (ECUs) without discretion.~\cite{currie2017hacking}.
    \item \textbf{OCPP}: Nasr et al.~\cite{nasr2022power} reported 13 types of vulnerabilities (e.g., missing authentication, hard-coded credentials, and missing rate limit) in 16 real-world CSMSs. Those vulnerabilities can be further exploited to compromise the lower-level EVSE by embedding malware into patches, thus disrupting the charging process and manipulating the default setting of the EVSE.
    \item \textbf{USB port on EVSE}: Although potential vulnerabilities of USB ports were demonstrated in several studies~\cite{harnett2018doe,acharya2020public}, the first attack via the external interface, e.g., USB or serial interfaces, was reported by the Idaho National Laboratory~\cite{carlson2018cyber,rohde2019cyber}. After obtaining physical and remote access to the EVSE, researchers successfully manipulated the modular power electronics modules in EVSE ports equipped with J1772 CCS and CHAdeMO protocols, thus disrupting the charging process.
\end{itemize}

\subsection{Possible Attacks on ESMS}%%%Potential attack vector on ESMS
\begin{table*}[ht]
    \caption{Summary of possible attacks on the ESMS}
    \begin{tabular}{l|l|p{6.7cm}|l}\hline
          Type of attacks & Targeted assets & Impacts & Ref. \\\hline
        EV-EVSE interface tampering & EV-EVSE communication & Terminate charging unless manually reconnecting & \cite{kohler2022end}\\\hline
        Distributed Denial-of-Service & CSMS and its communication with EVSE & Unavailability of EVSE, delayed response from CSMS& \cite{antoun2020detailed}\\\hline
        Electronic control manipulation & EVSE Human Machine Interface (HMI) or SEV & Modify the HMI front panel display (SoC, time remaining), disrupt controls coordination between power modules & \cite{rohde2019cyber}\\\hline
        Man-in-the-Middle attack & Communication between EVSE, EV, and EV driver & Track
issues, payment fraud, spoofing for a free service & \cite{bhusal2021cybersecurity}
        \\\hline
        \end{tabular}
    \label{tab:summary_of_attacks}
\end{table*}

In the ESMS, the vulnerabilities identified during the charging process create a \textit{linked sequence} from the external USB port in the maintenance interface, via the connector, and to the BMS and ECUs in the SEV. These vulnerabilities provide various attack surfaces for different malicious attacks, including an EV-EVSE tampering attack, Distributed Denial-of-Service (DDoS) attack, Man-in-the-Middle (MitM) attack, and electronic control manipulation. We summarize these attack vectors and their corresponding impacts on the charging service in Table~\ref{tab:summary_of_attacks}.
Specifically, the EV-EVSE interface tampering is only possible by physically deploying the off-the-shelf radio near the EVSE, which will significantly suffer from high attacking costs for large-scale impacts. Moreover, the charging process will completely terminate unless manually reconnecting to the connector, making the attack easily detectable through manual reporting of malfunctions.
Similarly, the consequences of DDoS and electronic control manipulation are mainly explicit and easier to detect, e.g., delayed server response and web service disruption for hours~\cite{antonakakis2017understanding}, which have little impact on the ESMS in the long run.
As for the MitM attack, the primary target is the data privacy issue, yet few impacts have been reported on the system performance of ESMS. 
In summary, we note that the above attacks target either one-time breakdowns or data privacy issues, which do not align with our research goal. On the other hand, our proposed DCA serves as a special type of the FDIA that stealthily falsifies the SEV to accept a lower charging rate, leading to a delayed charging service and long-term degradation of the ESMS. In the following sections, we will present the stage-by-stage DCA development and the AD techniques for the DCA detection.

\section{DCA Development and Anomaly Detection}\label{sec:model}

In this section, we focus on the development of the DCA model and its evaluation against AD techniques. Specifically, we contribute to a novel DCA model that includes the attack vector, potential consequences for the ESMS, and implementation details. To evaluate the efficacy and robustness of the proposed DCA, we introduce five different AD techniques and incorporate them into a SIRS process, enabling the modeling of DCA dynamics in a more realistic and practical manner.

\subsection{DCA Modeling}\label{sec:dca_implementation}
The DCA is a distinct form of FDIA that aims to disrupt the charging service for the SEV fleet by injecting falsified SoC information into the SEV. The attacker exploits vulnerabilities in physical entries (e.g., USB ports) and communication interfaces and protocols (e.g., CAN bus and OCPP) to manipulate the SEV into accepting a reduced charging rate while still ensuring SEV battery's safety (e.g., avoiding excessive current or voltage). The DCA first compromises a set of EVSE via the USB port and targets the delay of charging services for the SEV fleet. This deviation from the normal operation can hardly be detected by SEV drivers and standard AD techniques due to the wide range of charging duration, from several minutes to 1-2 hours~\cite{lei2022understanding}. These minor delays in individual charging activities will result in local congestion at EVSEs and the unavailability of SEVs, eventually leading to a cascading failure in ESMS.

It is worth noting that a straightforward DCA can be executed by directly charging the SEV at a lower rate. However, the BMS in SEV continuously monitors the input current via the CAN bus and ECUs. The BMS will raise the alarm if the deviation from the optimal charging current surpasses a set threshold. Therefore, special care is required to devise a DCA model that can stealthily slow down the charging process.
We next outline the full steps of the DCA implementation as below (also illustrated in Fig.~\ref{fig:layout_DCA}, where the falsified data are highlighted as red arrows):

\begin{figure}[h]
\centering
    \includegraphics[width=0.98\linewidth]{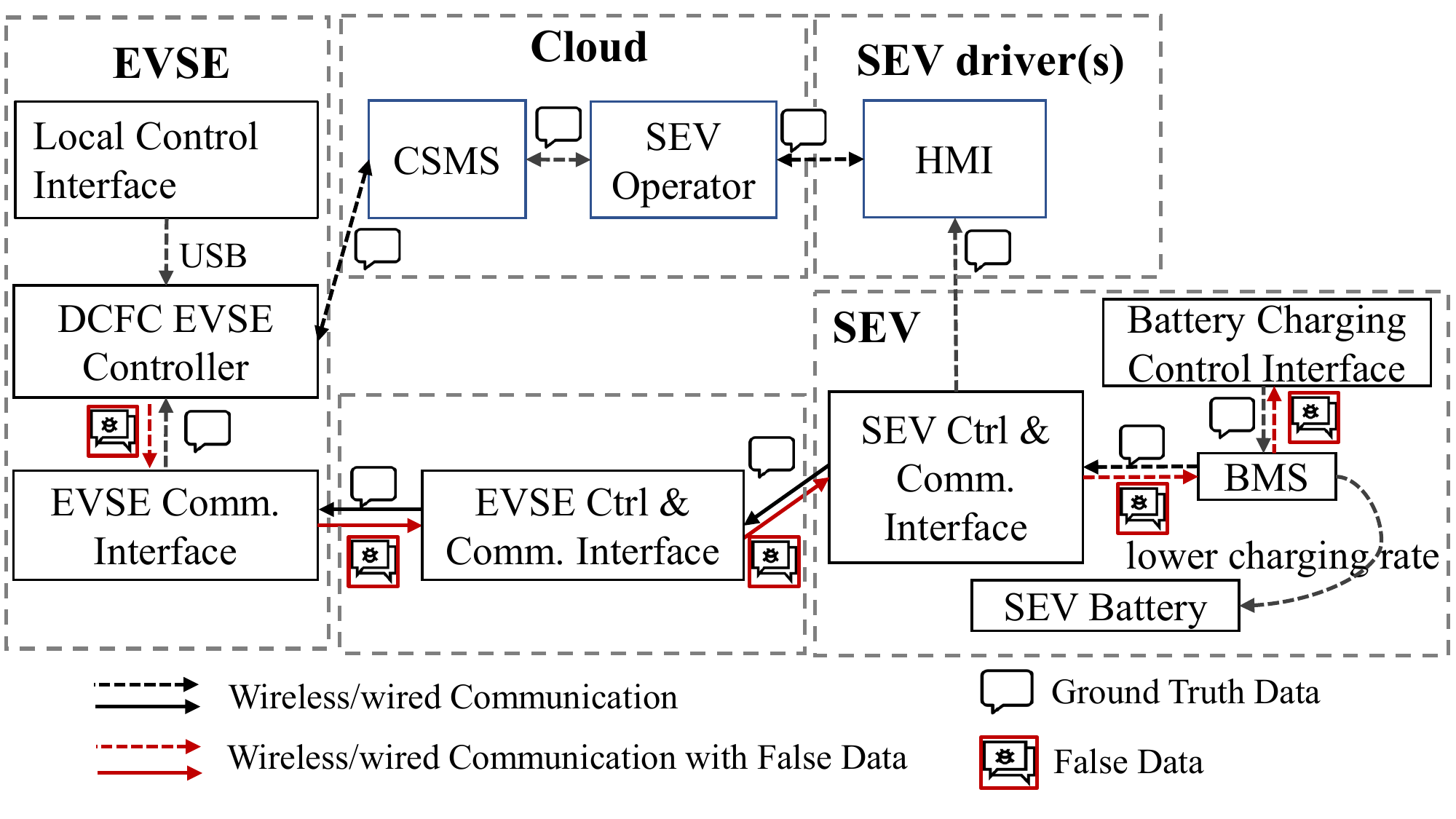}
    \caption{A block diagram of DCA}
    \label{fig:layout_DCA}
\end{figure}

\begin{enumerate}
    \item Compromise the EVSE via the USB port.
    \item Before the charging starts, the compromised EVSE sends true configuration data to the SEV for the compatibility check.
    \item During the charging process: 
    \begin{enumerate}
        \item The DCFC EVSE controller collects the ground-truth information of the SEV battery via CAN and reports it to the CSMS via OCPP.
        \item The DCFC EVSE controller sends falsified SoC information to the BMS in SEV through the EVSE control\&communication interface (CAN bus in CHAdeMO and PLC in CCS protocol).
    \end{enumerate}
\end{enumerate}

Step 3(b) serves as the core of our DCA for delaying the charging service. It ensures that the battery safety requirement, e.g., maximum current and voltage, are met while bypassing the BMS's monitoring for the optimal charging current. Furthermore, our DCA guarantees that the charging log information uploaded to the cloud (e.g., CSMS and SEV Operator) remains unchanged, allowing for successful cross-verification in the cloud.

\begin{figure}[h]
    \subfloat[]{\includegraphics[width=0.48\linewidth]{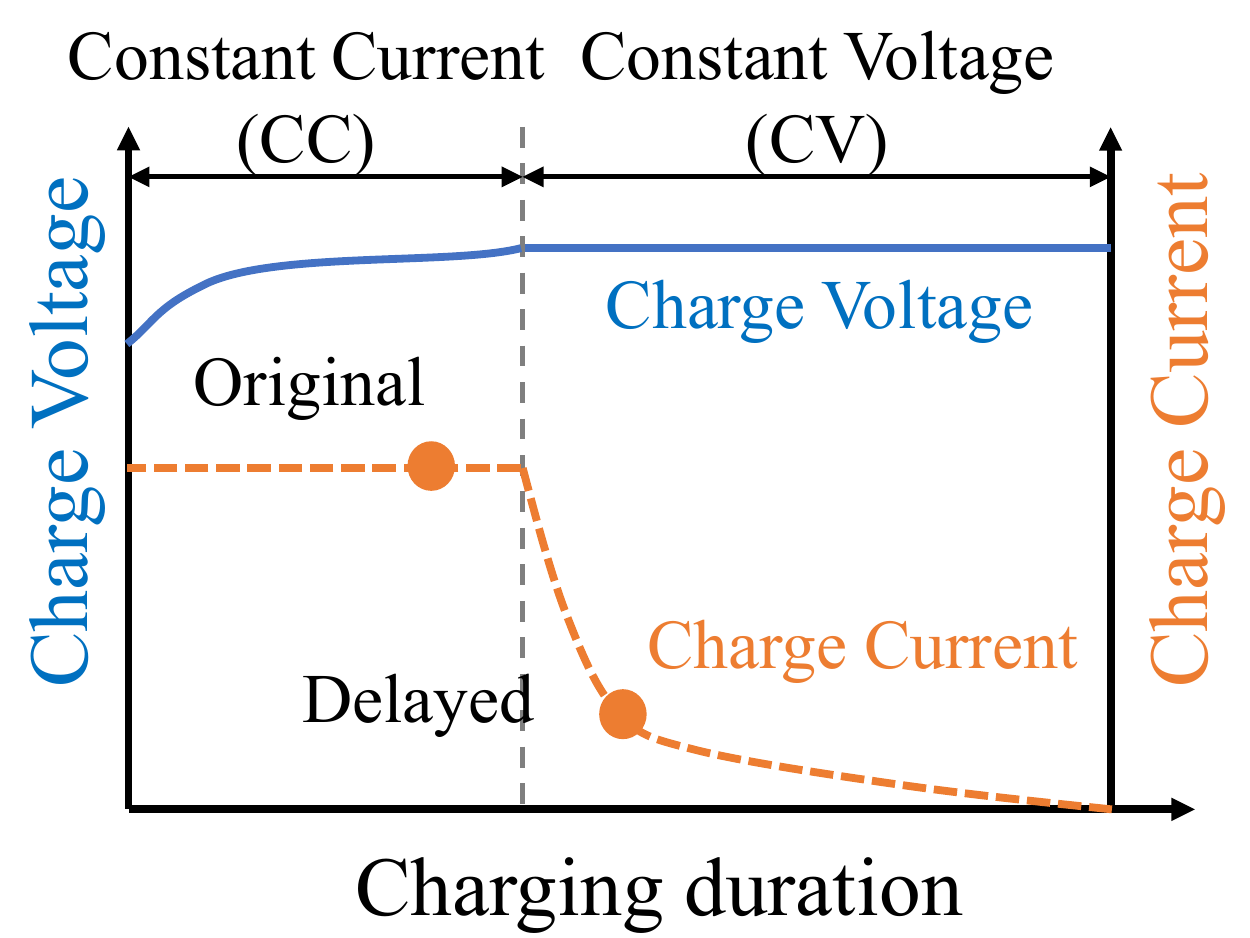}\label{fig:cccv}}
    \subfloat[]{\includegraphics[width=0.48\linewidth]{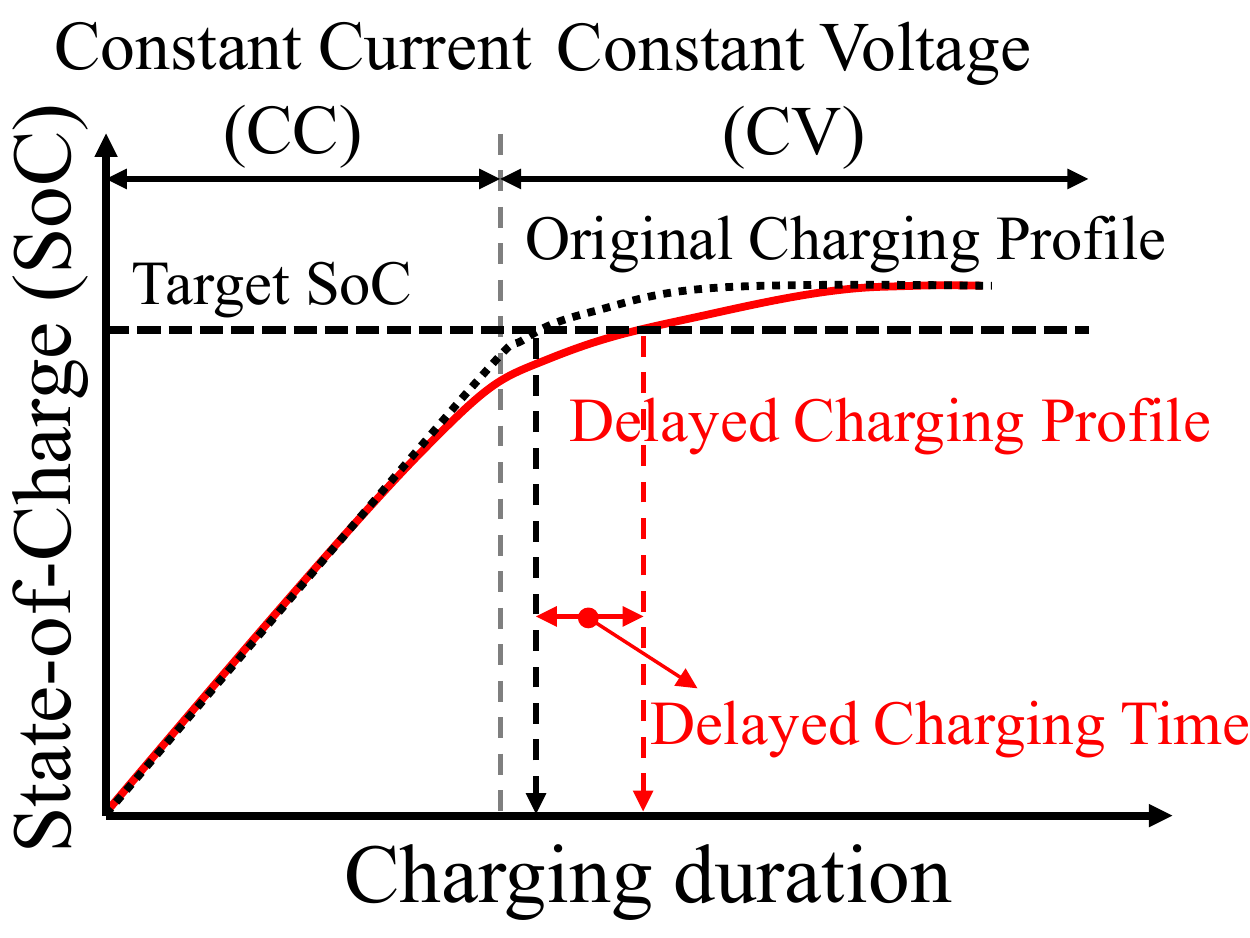}\label{fig:cccv_soc}}\\
    \caption{Charging profiles under normal operation and DCA: (a) CC-CV charging process and (b) SoC profile}
    \label{fig:layout_cccv}
\end{figure}

To better understand the SEV battery's recharging and monitoring mechanisms, we present the charging profiles of the normal operation and the DCA in Fig.~\ref{fig:layout_cccv}. We assume that the BMS adopts the constant-current-constant-voltage (CC-CV) recharging scheme that is widely used for Lithium-ion batteries. Under the proper design of falsified SoC information, the charging service can be delayed within safe limits.

The CC-CV recharging scheme, shown in Fig.\ref{fig:cccv}, maps the charging duration (or SoC information) and charging rate (i.e., charge voltage and current) in a one-to-one relationship. During the constant current (CC) phase, the charge voltage increases gradually to reach its maximum. At a certain SoC tipping point (e.g., 80\%), the charge current begins to decrease gradually to zero in the constant voltage (CV) phase. This relationship allows the BMS to detect a change in the charging rate if the SoC information and charge voltage (or current) do not match. In light of this, our DCA leverages this characteristic to manipulate the SoC reported to the BMS, leading to a reduction in the optimal charging current. The difference between the original and delayed charging currents are shown in Fig.~\ref{fig:cccv}, indicating that the optimal charging current under DCA will be less than or equal to the original charging rate. As illustrated in Fig.~\ref{fig:cccv_soc}, the lower charging rate will produce a smoother charging profile and lead to a longer charging duration to reach the target SoC, while still respecting the safety limits of charge voltage and current.

\subsection{DCA Dynamics in SIRS Process}

To model the DCA dynamics, we consider an SIRS process. First, the DCA is launched at a set of infectious EVSE ports. As discussed in Sec.~\ref{sec:dca_implementation}, the infected EVSE controller will send falsified SoC information to the SEV's BMS, resulting in a lower optimal charging current and an extended time to reach the same target SoC. As shown in Fig.~\ref{fig:SIRS_model}, the DCA model comprises three states of EVSE: \textbf{S} (Susceptible), \textbf{I} (Infectious), and \textbf{R} (Removed), as well as parameters $\beta$ and $\gamma$ that denote the transmission rate and recovery rate, respectively. In addition, we denote by $\tau$ the repair rate, where $1/\tau$ represents the repair time, indicating the time required for an EVSE to return to the susceptible state. The implementation is outlined in Algorithm~\ref{algo:cyberattack_simulation}.

\begin{figure}[h]
    \centering
    \begin{tikzpicture}[->,>=stealth',shorten >=1pt,auto,node distance=3cm,
                    thick,main node/.style={circle,draw,font=\sffamily\Large\bfseries}]
      \node[main node] (1) {S};
      \node[main node] (2) [right of=1] {I};
      \node[main node] (3) [right of=2] {R};
      \path[every node/.style={font=\sffamily\small}]
        (1) edge node [above]  {$\beta$} (2)
        (2) edge node [above]  {$\gamma$} (3)
        (3) edge [bend left] node [above] {$\tau$} (1);
    \end{tikzpicture}
    \caption{SIRS model under DCA}
    \label{fig:SIRS_model}
\end{figure}
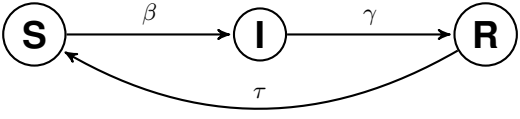
\begin{algorithm}[h]
    \caption{SIRS model in the ESMS}\label{algo:cyberattack_simulation}
    \footnotesize
	\begin{algorithmic}[1]
         \renewcommand{\algorithmicrequire}{\textbf{Input:}}
         \renewcommand{\algorithmicensure}{\textbf{Output:}}
         \REQUIRE At time step $t$, set of SEVs $\mathcal{V}^{t}$, set of EVSE $\mathcal{S}^{t}=\mathcal{S}^{t}_{S}\bigcup\mathcal{S}^{t}_{I}\bigcup\mathcal{S}^{t}_{R}$\\
         \ENSURE  Set of EVSE $\mathcal{S}^{t+1}$ with updated status.\\
    	    \FOR[*/loop all EVSE*/]{$i \in \mathcal{S}^{t}$}
                \IF{$i$ is susceptible}
                    \STATE \textbf{Calculate} $x_{i}^{S\rightarrow I}\leftarrow B(1,\beta)$ \{*/$B$ is Binomial distribution*/\}
                    \IF{$x_{i}^{S\rightarrow I}==1$}                
                        \STATE \textbf{Update} the status to infectious
                        \STATE $\mathcal{S}_{I}^{t+1}\leftarrow \mathcal{S}^{t}_{I}\bigcup\{i\}$; $\mathcal{S}_{S}^{t+1}\leftarrow \mathcal{S}^{t}_{S}\setminus\{i\}$
                        \STATE \textbf{Record} the time stamp $t_{i}^{I}\leftarrow t$
                    \ENDIF
                \ELSIF[*/Repair is finished*/]{$i$ is removed \AND $t-t_{i}^{R}\ge 1/\tau$}
                    \STATE \textbf{Update} the status back to susceptible
                    \STATE $\mathcal{S}_{S}^{t+1}\leftarrow \mathcal{S}_{S}^{t}\bigcup\{i\}$; $\mathcal{S}_{R}^{t+1}\leftarrow \mathcal{S}_{R}^{t}\setminus\{i\}$
                \ENDIF
            \ENDFOR
            \FOR[*/loop all infectious EVSE*/]{$i \in \mathcal{S}^{t}_{I}$}
                \STATE \textbf{Proceed} $\delta_{i}^{I\rightarrow R}\leftarrow$ {\textsc{AnomalyDetection}($d_{i},t_{i},c_{i}$)}
                \IF[*/EVSE is identified to be an anomaly*/]{$\delta_{i}^{I\rightarrow R}==1$}
                    \STATE \textbf{Conduct} alarm validation
                    \IF{alarm is true positive}
                        \STATE \textbf{Update} the status to removed.
                        \STATE \textbf{Record} the time stamp $t_{i}^{R}\leftarrow t$
                        \STATE $\mathcal{S}^{t+1}_{R}\leftarrow \mathcal{S}^{t}_{R}\bigcup\{i\}$; $\mathcal{S}_{I}^{t+1}\leftarrow \mathcal{S}_{I}^{I}\setminus\{i\}$
                    \ENDIF
                \ENDIF
            \ENDFOR
    \end{algorithmic}
\end{algorithm}

\subsubsection{S-I transmission}
In the S-I transmission process, an EVSE is compromised by either a wireless communication interface (e.g., between CSMS and EVSE under OCPP) or an external physical entry (e.g., USB port). For this study, we make a conservative assumption such that the \textit{EVSE is only infected by the physical access via the USB port}, as this approach has been validated to disrupt the SEV's charging service by the Idaho National Laboratory~\cite{carlson2018cyber,rohde2019cyber}.
We also note that the malware may also spread through the EVSE communication network and SEV-EVSE connection~\cite{mousavian2017risk}. These stronger assumptions may contribute to a higher transmission rate $\beta$. However, there is currently a lack of real-world evidence within the ESMS to support such assumptions. As such, these types of malware infection will be the focus of future research as more vulnerabilities in the ESMS are identified. 

\subsubsection{I-R transmission}
The I-R transmission relies on the AD algorithm based on the charging log information. Upon a positive alarm, the EVSE operator will first verify it through the EVSE communication network (assumed within one minute) and send out technician for repair service. If the alarm is confirmed to be a true positive (the EVSE is infectious), the EVSE will be isolated and the technicians will inspect and repair. Therefore, the recovery rate ($\gamma$) strongly depends on the effectiveness of the AD algorithm, which will be introduced in Section~\ref{sec:AD}.

\subsubsection{R-S transmission}
The R-S transmission occurs after the completion of the repair process for the infected EVSE. The duration of this process is determined based on the mean-time-to-repair (MTTR), denoted by $\tau$. During this period, the EVSE is out of service, and the queued SEV will relocate to other available EVSE ports.

\subsection{Anomaly Detection for DCA}\label{sec:AD}
To examine the robustness of the DCA model, we incorporate the AD algorithms in the ESMS that proactively monitor the system charging performances and protect against potential threats. The core purpose of the AD algorithms is to prevent the SEV fleet from experiencing delays in charging service, thereby mitigating the potential for cascading failures in the entire system due to the local congestion and excessive downtime of SEV supply, especially during the peak hour. 
In particular, we will focus on five AD techniques considering (1) the features of anomalies and (2) the ESMS operation. For the former, the anomalies in our study are assumed to be contextual~\cite{song2007conditional}. For instance, a short charging duration (e.g., \SI{20}{min}) may be considered anomalous during the nighttime but acceptable during the peak hours~\cite{lei2022understanding}. For the latter, we treat the historical charging performance as the baseline (e.g., training set), assuming that it only includes data instances collected during the normal operation before the DCA launched. 
We next introduce five AD techniques as follows:

\subsubsection{Isolation Forest (IF)}
The IF algorithm was first proposed by Liu et al.~\cite{liu2008isolation} for the purpose of AD and was later used for the purpose of cyberattack detection~\cite{elnour2021application}.
The IF-based detection algorithm proceeds as follows. We first train the IF model using the benign historical charging log data to understand the properties of the normal operation. The charging log data consists of the charging duration, time of day, and the initial SoC, denoted by $(d_i,t_i,c_i)$. With a pre-defined false alarm rate $\alpha_{I}$ and the number of estimators $n$, we are able to train an IF model consisting of $n$ proper binary trees, where $\alpha_{I}$ of the samples are considered as the anomaly. 
% A higher $\alpha_{I}$ denotes a more sensitive detection, which may identify more anomalies and require more effort for verifying the alarms.
% {\color{magenta}detection cost [what does the detection cost refer to? is it refer to high information loss or refer to inference time? or other?]}
% To construct the isolated tree, a binary splitting procedure is recursively conducted by randomly selecting an attribute and a split value, which terminates until all samples are isolated or the tree reaches a height limit.
Next, we collect batches of charging log data online as testing data. We denote the anomaly score of sample $i$ in the testing set by $s_{i}\in[0,1]$, which takes the form:

\begin{equation}
    s_{i} = 2^{-\frac{\Bar{h}_{i}}{\Bar{h}}}
\end{equation}
where $\Bar{h}_{i}$ denotes the average path length of a sample $i$ from a collection of isolation trees, and $\Bar{h}$ is the average path length of an unsuccessful search, which is adapted to normalize the $\Bar{h}_{i}$. Thus the anomaly score can be understood as the efforts to find a path in the isolated tree, where the anomalies can be identified at the early stage of exploration.
% To better understand the property of anomaly score, we list the following three cases: 
% \begin{itemize}[leftmargin=*]
%     \item $s_{j}\rightarrow0$: the anomaly samples are likely to be spitted within a lower depth of the tree.
%     \item $s_{j}\rightarrow1$: the sample can be regarded as normal when the depth of search is far longer than the average depth $\Bar{h}$. 
%     \item $s_{j}\rightarrow 0.5$: no distinct anomaly is detected, such that $\Bar{h}_{j}\rightarrow{\Bar{h}}$.
% \end{itemize}
By further incorporating with a false alarm rate $\alpha_{I}$, the anomaly is defined by a binary indicator $y_{i}$, where $y_{i}=1$ if $s_{i}< \alpha_{I}$ (the sample $i$ is identified to be an anomaly) and $0$ otherwise.

\subsubsection{Kullback–Leibler Divergence (KLD)} 
The KLD~\cite{kullback1951information} measures the differences between two probability distributions $p(x)$ and $q(x)$ regarding the event $x\in\mathcal{X}$, denoted by $D(p||q)$, which can be expressed as:

\begin{equation}
    D(p||q) = \sum\limits_{x\in\mathcal{X}} p(x) \log{\frac{p(x)}{q(x)}}
\end{equation}
where the $p(x)$ and $q(x)$ are assumed to represent the historical distributions of one EVSE under normal operation and DCA, respectively. 
Specifically, $p(x)$ is obtained based on the historical charging log data, and $q(x)$ represents the real-time charging logs. For the charging log sample $x$, each log can be expressed as a triple $(d_i,t_i,c_i)$. 
% Let $\alpha_{KLD}$ be the sensitivity level of the KLD-based AD technique.
And the anomaly is identified if the samples are associated with the KLD $D(p||q)$ larger than a pre-defined threshold $\alpha_{KLD}$. 
% Note that a cross-validation procedure is conducted to find the range of thresholds that yields a good F1 score.

\subsubsection{K-Means clustering (KMeans)} The KMeans clustering method~\cite{hartigan1979algorithm} is a cluster-based algorithm that groups the input samples into $K$ disjoint clusters based on the sample features. Specifically, we first obtain $K$ clusters based on the historical charging log data $(d_i,t_i,c_i)$. Next, we measure the distance between the online charging logs $(d_j,t_j,c_j)$ and the nearest centroid of the $K$ clusters. Given a sensitivity level of $\alpha_{KM}$, we report the EVSE $i\in\mathcal{S}$ as an anomaly if more than $\alpha_{KM}\%$ of the real-time samples are associated with a distance larger than the pre-defined threshold $D_{KM}$.

\subsubsection{Gaussian Mixture Model (GMM)}
The GMM is a model consisting of $K$ separate multivariate normal distributions $\mathcal{N}(\cdot)$, known as mixture components. Each mixture component is parameterized by $\theta_{k}:=\{\mu_{k},\Sigma_{k}\}$, where $\mu_{k}$ and $\Sigma_{k}$ represent the mean value vector and covariance matrix of the $k$-th mixture component. Let $\pi_{k}\ge 0$ be the associated weight for the $k$-th mixture component, where $\sum_{k=1}^{K}\pi_{k}=1$. We consider the probability density function for the GMM as a weighted sum of the mixture components:

\begin{equation}
    p(\mathbf{x}|\theta) = \sum\limits_{k=1}^{K}\pi_{k}\mathcal{N}(\mathbf{x}|\mu_{k},\Sigma_{k})
\end{equation}
where $\mathbf{x}$ is a collection of tuple $(d_i,t_i,c_i)$ with the dimension $K=3$. For the GMM-based AD technique, we define the anomaly as the samples $x_{i}$ with $p(x_{i}|\theta)<c_{G}$, where $c_{G}$ denotes the significance level of the GMM. An anomaly is detected if more than $\alpha_{G}\%$ samples for an EVSE are identified as an outlier.
Note that there is no analytical solution for the parameters $\theta:=\{\theta_k:k=1,\cdots, K\}$. Instead, we can estimate the parameters $\theta$ using the Expectation-Maximization (EM) algorithm~\cite{dempster1977maximum}. The EM algorithm proceeds by initiating a random set of parameters $\Hat{\theta}$ and iteratively estimating the optimal set $\Hat{\theta}^{\star}$ that can maximize the average log-likelihood given the training set $\mathbf{x}$. For the detailed implementation of the EM algorithm, we refer interested readers to Reynolds~\cite{reynolds2009gaussian}.

\subsubsection{Principal component classifier (PCC)}
The PCC~\cite{shyu2003novel} first obtains the principal components from the covariance matrix of the training data (assumed as the historical charging log data under normal operation). Next, an anomaly score is assigned to each online charging event based on its deviation from the principal components.
The anomaly score incorporates the \textit{major} and \textit{minor} components. 
Specifically, the former detects extreme observations (charging events with large variances). The minor components help to detect the values that are not outliers but inconsistent with the correlation structure as the normal operation. 
For each EVSE $i\in\mathcal{S}$, we define the EVSE $i$ as an anomaly if over $\alpha_{P}\%$ of the batch of real-time charging events are associated with a significant level of $c_{P}$ for either the major or minor components.

\section{Agent-based Simulation Platform}~\label{sec:simulator}
We develop an agent-based simulator~\footnote{\href{https://github.com/sguo28/DCA_Simulator}{https://github.com/sguo28/DCA\_Simulator}} to characterize the interactions between agents (SEVs) and the environment (passenger demand and EVSE ports). Our simulator has five components: matching, dispatching, repositioning, charging, and DCA unit. Specifically, the DCA unit includes the SIRS-based DCA model and AD techniques to demonstrate the robustness of DCA. In addition, we also design utility-based heuristic algorithms that guide unoccupied SEVs to the under-supply areas and match the SEVs to available EVSE. Key features in our simulator are listed below (for more detailed settings, see Qian et al.~\cite{qian2022drop}).

\subsection{DCA Unit}

\subsubsection{SIRS-based DCA Model} We first assume all EVSEs are infectious after the warm-up period. The SEVs using the infectious EVSE ports will encounter a delayed charging service following the Gaussian noise $\Delta d \sim \mathcal{N}\left(\mu_{d}, \sigma_{d}\right)$ such that the charging duration is $d_i\leftarrow d_i+\Delta d$. If the infectious EVSE is detected, it will undergo the Infectious-Removed-Susceptible process, such that the EVSE can be infected again after being repaired.

\subsubsection{AD Techniques for DCA Detection} The infectious EVSE can be identified as an anomaly by comparing the real-time and historical charging log information (e.g., charging duration, time of day, and initial SoC), see details in Section~\ref{sec:AD} and Algorithm~\ref{algo:cyberattack_simulation}.
\subsection{Charging}
\subsubsection{Matching with EVSE Ports} We consider a utility-based heuristic such that each SEV is assigned to the EVSE with the highest utility scaled by a soft-max function. For each EVSE port $j\in\mathcal{S}$, a shorter queuing time $q_{j}$ and travel time $t_{\cdot j}$ will lead to a higher utility. Specifically, for an SEV $i$, the probability of selecting EVSE $j$, $P^{c}(j|i)$, is shown below:
\begin{equation}
    P^{c}(j|i)=\frac{\exp{\left(-q_{j}t_{ij}\right)}}{\sum\limits_{j\in\mathcal{S}}\exp{\left(-q_{j}t_{ij}\right)}}\label{eq:evse_matching}
\end{equation}
where $\sum\limits_{j\in\mathcal{S}}P^{c}(j|i)=1$ and $P^{c}(j|i)>0$ for each location $i$.
\subsubsection{Charging Service and Queuing} 
The charging service follows the first-come-first-serve rule.

\subsubsection{SIRS Process} 
Only the EVSE in susceptible and infectious statuses can serve SEVs. If an infectious EVSE is detected for repair, the SEV in the queue will relocate to a nearby EVSE with the highest utility following Eq.~\eqref{eq:evse_matching}.

\subsection{Repositioning} 
To better describe the status-quo scenario~\cite{millard2021ridehail}, we assume that the idled SEVs will reposition to an under-supplying area. Specifically, we conduct a utility-based heuristic analogous to the EVSE matching (see Eq.~\eqref{eq:evse_matching}). The utility score is calculated based on the supply-demand gap and travel time. At location $j$, the supply-demand gap is defined by the gap between the number of fulfilled orders ($N_{j}^{\text{order}}$) and idled SEVs ($N_{j}^{\text{idle}}$) in the past \SI{15}{min}.
For an empty SEV at location $i$, the probability of selecting location $j\in\mathcal{Z}$, $P^{r}(j|i)$, takes the form of a soft-max function, shown as follows.

\begin{equation}
    P^{r}(j|i)=\frac{\exp{\left(\frac{N_{j}^{\text{order}}-N_{j}^{\text{idle}}}{t_{ij}}\right)}}{\sum\limits_{j\in\mathcal{Z}}\exp{\left(\frac{N_{j}^{\text{order}}-N_{j}^{\text{idle}}}{t_{ij}}\right)}}
\end{equation}
where $\mathcal{Z}$ denotes the set of hexagonal zones, and $t_{ij}$ is the travel time between locations $i$ and $j$. 
\subsection{Matching with Requests} 
We conduct a greedy matching strategy, which sequentially assigns idle vehicles to the open orders to achieve the shortest travel time within the permissible waiting time threshold. These include the idling and cruising SEVs that are traveling to the relocation destination. For the cruising SEVs, it is considered available for matching at its real-time location. After being matched, it will stop cruising and move to the assigned pick-up location.
\subsection{Dispatching} The SEVs with assigned requests will be dispatched to the pick-up location. The idled SEV will, on the other hand, be guided to the target location with the highest utility considering the supply-demand pattern and travel time. The route is generated and processed from the Open Source Routing Machine (OSRM) engine~\cite{luxen-vetter-2011}, which provides the detailed real-time location for each simulation tick.
% \noindent \textbf{Order generation}: The arrival location and time of passengers and their destinations strictly follow the historical trip data as the input. Each passenger is associated with a maximum waiting time beyond which they will reject further services. 

\section{Numerical Experiments}\label{sec:experiments}
\subsection{Case Study Area}
We conduct a real-world case study in NYC to demonstrate the effectiveness of the DCA model on the ESMS. In particular, we develop a high-fidelity simulation platform extended from our previous study~\cite{qian2022drop}. The simulation environment is updated every minute and covers 1,347 hexagon cells (each covering an area of 0.14 square miles).
We use NYC taxi data in May 2016 as the input~\cite{nyc_taxidata2016}, and all trips are assumed to originate and head to the centroid of the hexagon cells. In addition, we obtain the real-world locations of EVSE ports from the Alternative Fuels Data Center~\cite{afdc2022charginglocation}. Note that we only consider the DCFC ports considering the system efficiency and poor accessibility for taxi drivers to home charging in NYC~\cite{moniot2022estimating}. The travel time between each pair of centroids is obtained by overlaying with the actual road network and then querying from the OSRM engine. 

We perform a downscaled status-quo experiment of the taxi system by randomly sampling 25\% of the historical trip record (around 130,000 trips). As existing EVSE facilities can hardly support the charging demand of the 100\% electrified SEV fleet~\cite{moniot2022estimating}, we also consider enlarging the number of charging piles of the real-world charging stations~\cite{afdc2022charginglocation}.
To find the best combination of the SEV fleet size, demand, and the number of EVSE ports, we conduct a cross-validation procedure by first fixing 25\% of the taxi trip demand and tuning the fleet size from 1,300 to 1,500, incremented by 50 and increasing the number of EVSE by 1.5 to 3.0 times stepped by 0.5. 
The best combination consists of 1,400 SEVs and 215 DC Fast EVSE ports to serve 25\% of real-world daily taxi trips. The baseline scenario is justified by real-world evidence (see Figs.~\ref{fig:occupancy_basic}-\ref{fig:evse_occupancy_basic})
Unless otherwise specified, available SEVs will be dispatched to passengers that can be reached within \SI{20}{min}, and the maximum waiting time for passengers is \SI{10}{min}. Finally, we consider a 4-week simulation using three random seeds, where the first week is regarded as a warm-up, and the other three weeks are the validation period. The DCA is launched at the end of the warm-up.

\subsection{Parameter Setting}

We summarize in Table~\ref{tab:params_description} the parameter setting in this study, including SEV, mobility service, and the SIRS process.
\begin{table}[h]
\footnotesize
    \caption{Parameter assumption}
    \begin{tabular}{p{2.35cm}|p{4.15cm}|l}\hline
        Item                 & Assumptions & Ref. \\
        \hline\multicolumn{3}{l}{\textbf{{SEV}}}\\\hline
        SEV prototype &2019 Tesla Model 3 Standard Range&\cite{doe2022fuel} \\
        Charging profile    &  Constant charging rate under an SoC below 80\%   &\cite{zheng2020systematic,Moloughney2021how} \\
        Target SoC & $\mathcal{N}(0.78,0.02)$ bounded by 0.80 & - \\
        Charging rate           & \SI{5.13}{mile/min} &  \cite{evdata2022charging} \\
        Battery capacity & \SI{50}{kWh} for \SI{220}{miles} & \cite{evdata2022charging,doe2022fuel}\\
        \hline\multicolumn{3}{l}{\textbf{{Mobility service}}}\\\hline
        Payment  & $\$0.631\times \Delta \text{dist} + \$0.287\times \Delta \text{time}$ & \cite{qian2022drop,nyc_driver2016}\\
        \hline\multicolumn{3}{l}{\textbf{{SIRS process}}}\\\hline
        Transmission rate $\beta$              & 0.1       & -   \\
        Recovery rate $\gamma$ & Related to AD techniques& Sec.~\ref{sec:AD}\\
        Repair rate $\tau$ &1/\SI{3}{hours} & \cite{farley2019avista} \\\hline%\bottomrule
    \end{tabular}
    \label{tab:params_description}
\end{table}

Specifically, we adopt the prototype of the Tesla Model 3 for the electric mobility service~\cite{lambert2021tesla}, which is supported by both Tesla Supercharging and CCS EVSE ports. Based on the field experiment by Moloughney~\cite{Moloughney2021how}, it took \SI{32}{min} for a Tesla Model 3 to reach $80\%$ SoC and another \SI{31}{min} for a full charge, where the charging rate for the first $80\%$ is observed to be approximately linear. Such a linear charging profile is also validated by the in-vehicular database~\cite{evdata2022charging}, which reports an average charging rate of
\SI{3.3}{\%SoC/min} (\SI{5.13}{miles/min}) for our prototype.
For the DCA model, a higher sensitivity level ($\alpha_{(\cdot)}$) results in a more sensitive AD, which may effectively identify the anomaly but incur a higher repair cost (e.g., sending out technician labor to inspect and repair). To compromise the EVSE, we make a conservative assumption based on real-world evidence~\cite{carlson2018cyber}. The susceptible EVSE is infected by manually inserting a USB drive every \SI{30}{min} with a transmission rate of $\beta=0.1$. The AD algorithm is conducted every \SI{30}{min}, and the technicians for repair service are sent out for on-site assistance every $\SI{30}{min}$. The whole repair process for an EVSE port is set as $\tau=$\SI{3}{hours}. According to the Avista EVSE pilot report~\cite{farley2019avista}, we assume that the MTTR is 15 days to address the issue completely and the average cost to repair is \$214 (including the warranty and non-warranty labor and material costs). We note that multiple times of on-site assistance (e.g., power cycling, inspection, repair, or replacement) are required to fully resolve the issue~\cite{farley2019avista}. Therefore, the repair cost is estimated to be \$1.78 per time.
\subsection{Simulation Scenarios}
We present the detailed scenario design considering the different time delays for the charging service from \SI{5}{min} to \SI{15}{min} and the sensitivity levels of the AD algorithms ($\alpha_{(\cdot)}$). 
% The scenarios are summarized in Table~\ref{tab:scenario_design}. 
We will consider three types of experiments: (1) baseline, (2) DCA without AD, and (3) DCA with AD. The attack-free baseline scenario is to justify the downscaled status quo under normal operation. We also compare the scenarios under DCA with and without the AD to explore the trade-off between repair cost and improvement of the system performance. Finally, we conduct sensitivity analyses on the AD techniques. The range of $\alpha_{(\cdot)}$ is determined by a cross-validation procedure, where the parameter settings are shown in Table~\ref{tab:cv_for_AD}.

\begin{table}[h]
\footnotesize
    \centering
    \caption{Cross-validation for AD techniques}
    \label{tab:cv_for_AD}
    \begin{tabular}{c|c|c}\hline
       AD  & Range of $\alpha_{(\cdot)}$ & Hyperparameters \\\hline
       IF ($\alpha_{I}$)  & 0.05 to 0.15 by 0.025 & - \\
       KLD ($\alpha_{KLD}$) & 1 to 5 step by 1 & - \\
       KMeans ($\alpha_{KM}$)  & 0.2 to 0.6 step by 0.1 &$D_{KM}=2.5$\\
       GMM ($\alpha_{G}$)& 0.05 to 0.55 step by 0.1 & $c_{G}=0.01$\\
       PCC ($\alpha_{P}$) & 0.3 to 0.8 step by 0.1 & $c_{P}=0.005$\\\hline
    \end{tabular}
\end{table}

\section{Results}\label{sec:results}
\subsection{Baseline Scenario}
\begin{figure}[h]
    \subfloat[SEV occupancy rate]{\includegraphics[width=0.5\linewidth]{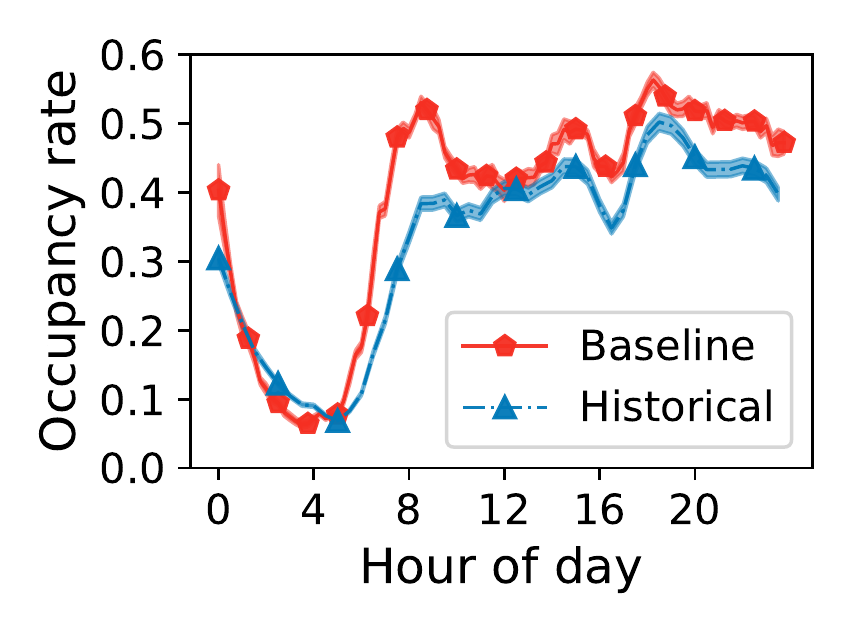}\label{fig:occupancy_basic}}\hfill 
    \subfloat[Order fulfillment rate]{\includegraphics[width=0.5\linewidth]{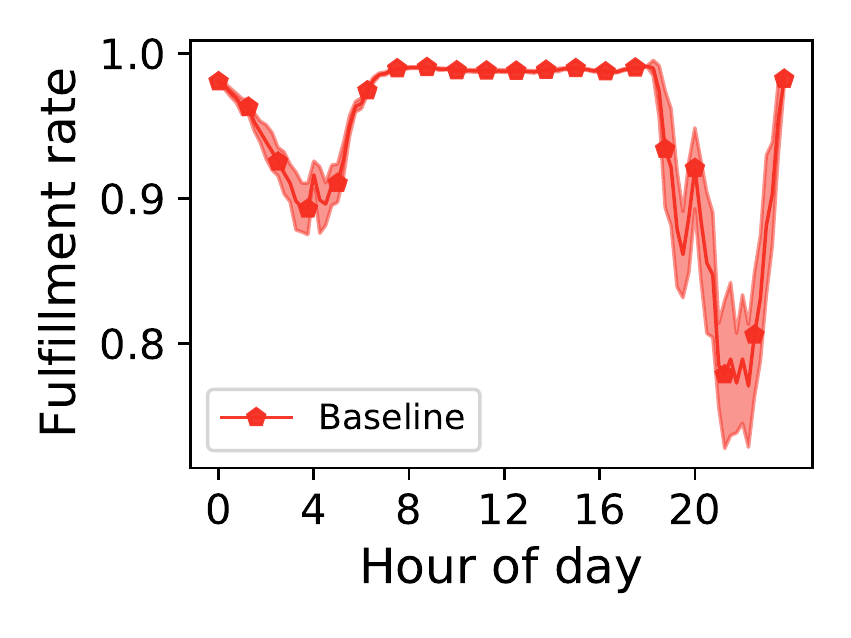}\label{fig:fulfillment_basic}}\vfill
    \subfloat[EVSE occupancy]{\includegraphics[width=0.5\linewidth]{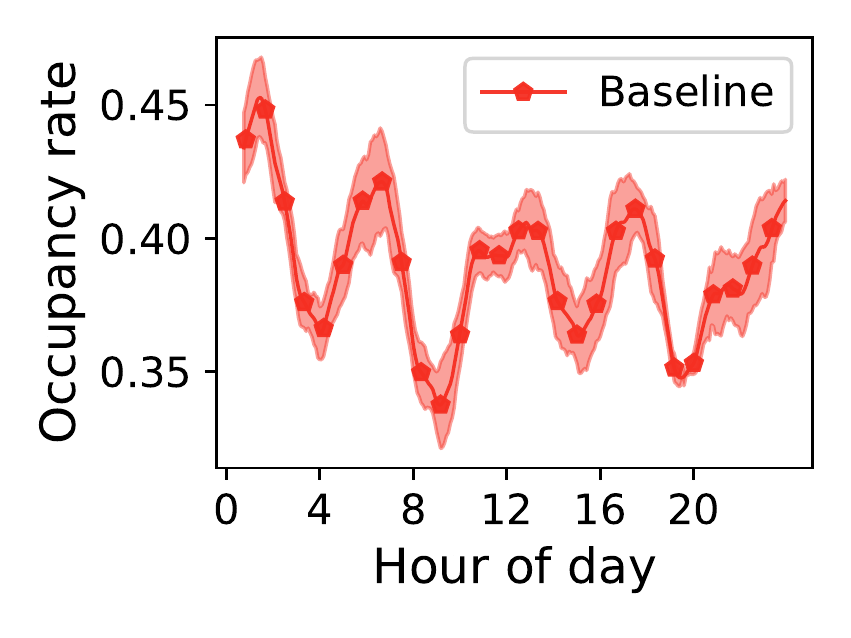}\label{fig:evse_occupancy_basic}}\hfill
    \subfloat[Queuing time]{\includegraphics[width=0.5\linewidth]{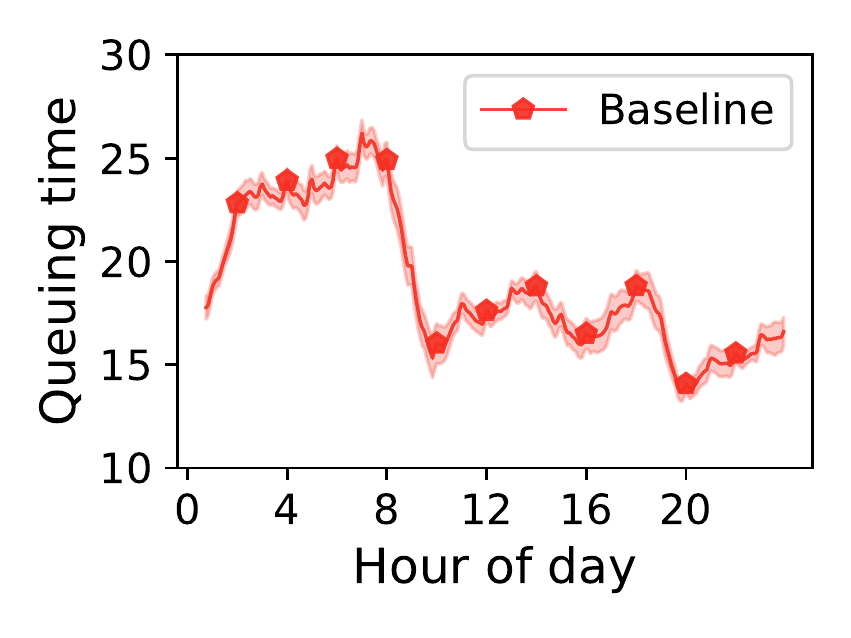}\label{fig:queuing_time_basic}}
\caption{\added{Major metrics in the baseline scenario}}~\label{fig:baseline_dynamics}
\end{figure}

This subsection presents the dynamics in the baseline scenario using real-world data. As seen in Figs.~\ref{fig:occupancy_basic}-\ref{fig:queuing_time_basic}, we show four primary metrics to justify the status quo simulation experiment, including (1) SEV occupancy rate, (2) order fulfillment rate, (3) EVSE occupancy rate, and (4) SEV queuing time at EVSE.
Fig.~\ref{fig:occupancy_basic} compares the SEV's occupancy rate with the historical NYC taxi fleet~\cite{nyc_taxifact2016}. The simulation results are observed to agree with the historical records, where the overall occupancy rate varies from less than 0.1 during 3 - 6 AM to nearly 50\% during the evening peak. The slightly higher occupancy rates in the simulation may result from the undersupply of the SEV fleet. In addition, we report an admissible fulfillment rate of nearly 100\% during the daytime and over 75\% during the evening peak (6 - 11 PM). Moreover, the validity of the EVSE occupancy rate and the SEV queuing time is confirmed by the real-world practice of the fully electrified taxi fleet in Shenzhen, China. In this case, an EVSE occupancy rate from 10\% to 80\%~\cite{wang2019sharedcharging} and a queuing time of \SI{10}{min} to \SI{40}{min}~\cite{dong2017rec} are considered to be plausible. Despite the different taxi market settings between NYC and Shenzhen, we believe that the charging dynamics in our simulation platform are permissible in the real-world ESMS, which can sufficiently justify our baseline scenario.

\subsection{System Performance of EVSE}
\begin{figure}[h]
    \subfloat[SEV queuing time at EVSE (min)]{\includegraphics[width=0.42\linewidth]{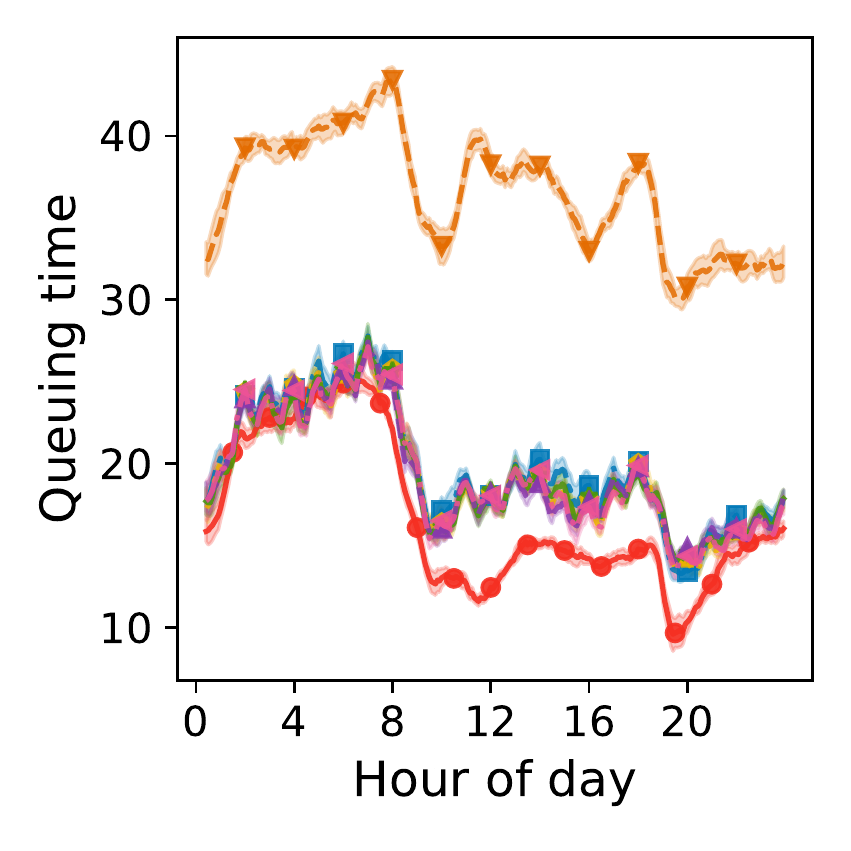}\label{fig:evse_queuing_time}}\hfill
    \subfloat[SEV charging time (min)]{\includegraphics[width=0.58\linewidth]{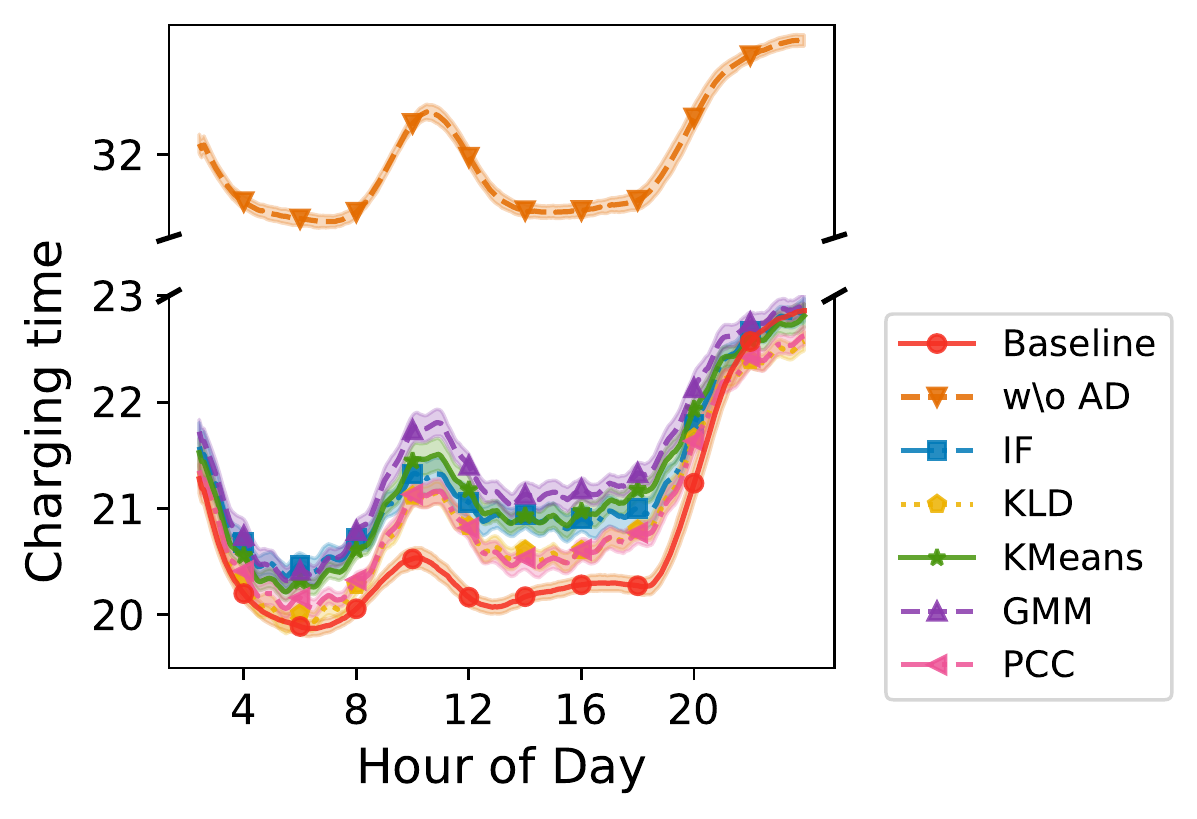}\label{fig:evse_charging_time}}
\caption{\added{Charging dynamics under the baseline scenario and \SI{10}{min} delay}}~\label{fig:baseline_dynamics_de}
\end{figure}

We next present the daily-average charging dynamics over the three-week period under the DCA, including queuing time at EVSE and charging duration, shown in Figs.~\ref{fig:evse_queuing_time}-\ref{fig:evse_charging_time}.

Specifically, Fig.~\ref{fig:evse_queuing_time} shows the queuing time of a SEV at an EVSE port. We will consider three cases: baseline, the DCA with and without AD under a fixed delayed charging time of \SI{10}{min}. For the baseline scenario, we observe two peaks during the early morning period (e.g., 6 - 8 AM) and the late afternoon period (around 4 PM), with the longest queuing time exceeding \SI{22}{min} compared with the shortest queuing time of about \SI{10}{min}. After launching DCA with a \SI{10}{min} delay, an additional queuing time of over \SI{15}{min} is observed during the noon period and even nearly \SI{20}{min} during the early morning period (e.g., 7-8 AM). The extended queuing time is due to the cascading effects. For instance, one SEV may relocate to the EVSE with the higher utility (e.g., average queuing time and travel time) after exceeding a queuing time threshold, which accelerates the snowballing of the excessive queuing time. With the AD techniques, the queuing time is greatly reduced by over \SI{10}{min}. However, there still exists space for improvement during the daytime, especially during the daytime (9 AM - 5 PM).
This is because there exists a proportion of EVSE that is removed for repair and out-of-service for at least $\tau=3$hours. In this regard, the SEVs in the queue have to relocate to other EVSE ports, resulting in local congestion at the EVSE ports in S and I statuses. As seen in Fig.~\ref{fig:evse_charging_time}, the charging duration can be reduced from over \SI{30}{min} to relatively the same levels as the baseline under all five AD techniques. Compared with the early morning and evening periods, the average delays in charging service are observed to be longer at around 6 AM and 2 PM, potentially due to a higher proportion of infectious EVSE ports.

\subsection{System Performance for SEV Fleet}
This subsection shows the SEV driver's weekly revenue loss under different lengths of charging delay with and without the AD models. To better understand the degradation of the mobility service, we also present two major dynamics, including order fulfillment rate and SEV occupancy rate. 

\begin{figure}[h]
    \centering
    \includegraphics[width=0.85\linewidth]{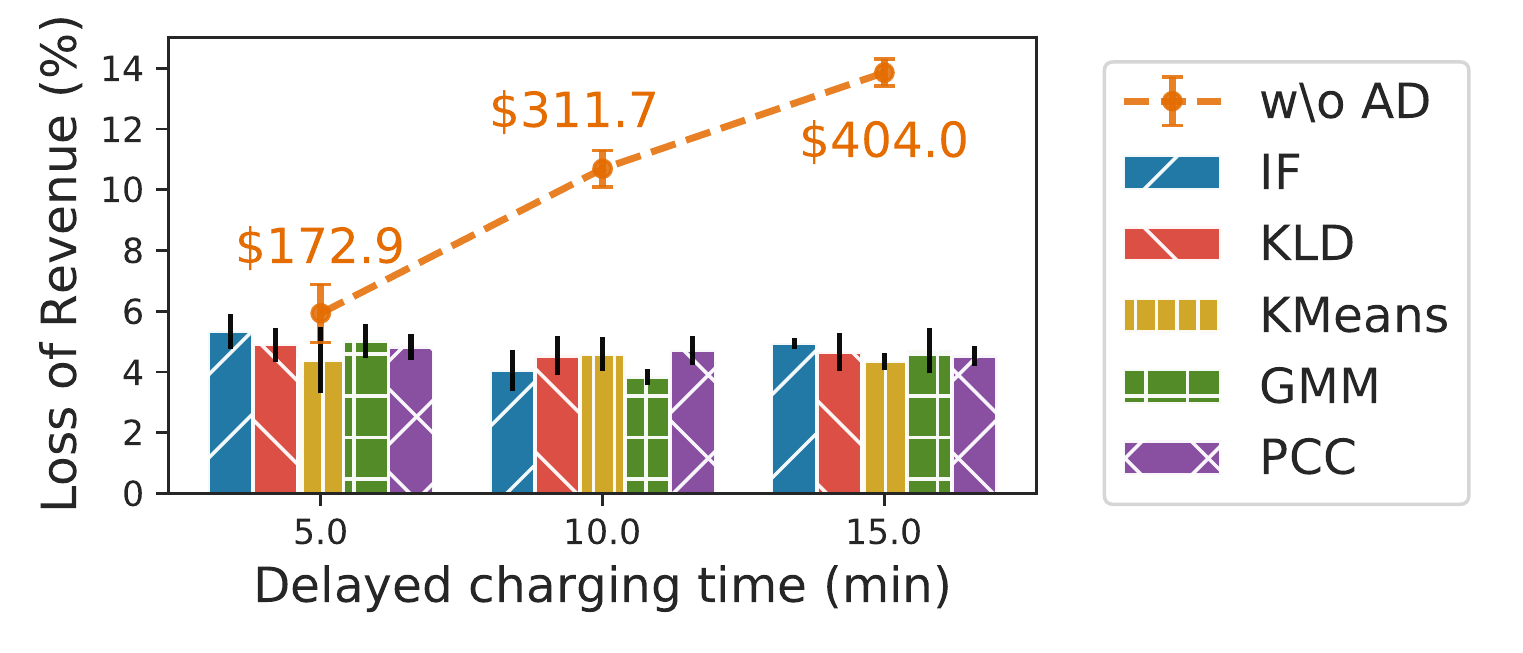}
    \caption{Comparison of total revenue during the last (4th) week in simulation}\label{fig:tot_revenue}
\end{figure}
Fig.~\ref{fig:tot_revenue} displays the system revenue loss during the last week in our simulation to capture the long-term impact of the DCA. One immediate observation is the strong linear relationship between delayed charging time and revenue loss. Specifically, without the AD, we report a system revenue loss of at least \$172.9 (5.9\%), \$311.7 (10.7\%), and up to \$404.0 (13.9\%) under the delay of 5, 10, and \SI{15}{min}, respectively. With a delay of \SI{15}{min}, the system revenue loss can be reduced to about \$140.1 (4.8\%) comparing all five AD techniques. Furthermore, our results show that for both IF and GMM, the revenue losses initially decrease when the delay in charging time is between 5 to 10 minutes, but start to increase as the delay grows from 10 to 15 minutes. This suggests that as the extended charging time increases, it may result in more severe charging delays at the system level, while also making it more susceptible to detection as the charging duration deviates further from the normal operation. This highlights the importance of carefully balancing between DCA's efficacy and robustness. 
Finally, despite the improvement in a system revenue loss of up to 8.6\% (from 13.9\% to 5.9\%), the underlined repair cost may grow exponentially with longer delay of charging, which will be evidenced in Fig.~\ref{fig:trade_off_revenue_detection}.

\begin{figure}[h]
    \subfloat[]{\includegraphics[width=0.98\linewidth]{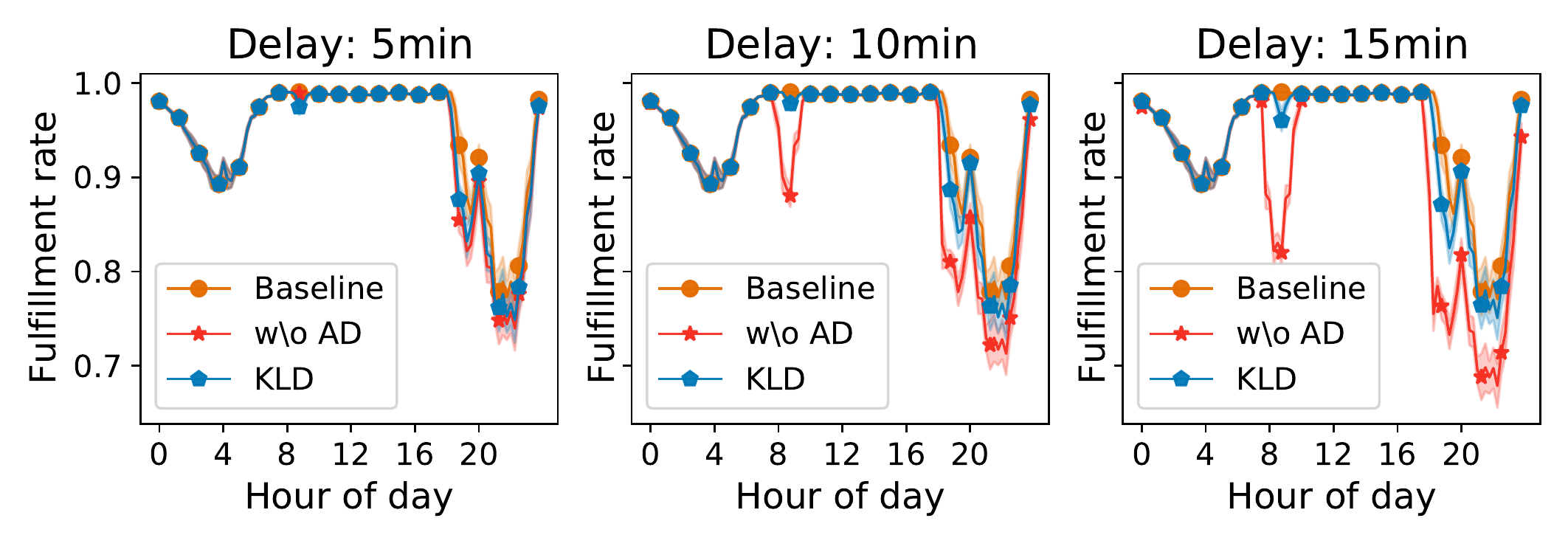}\label{fig:fulfillment}}\hfill 
    \subfloat[]{\includegraphics[width=0.98\linewidth]{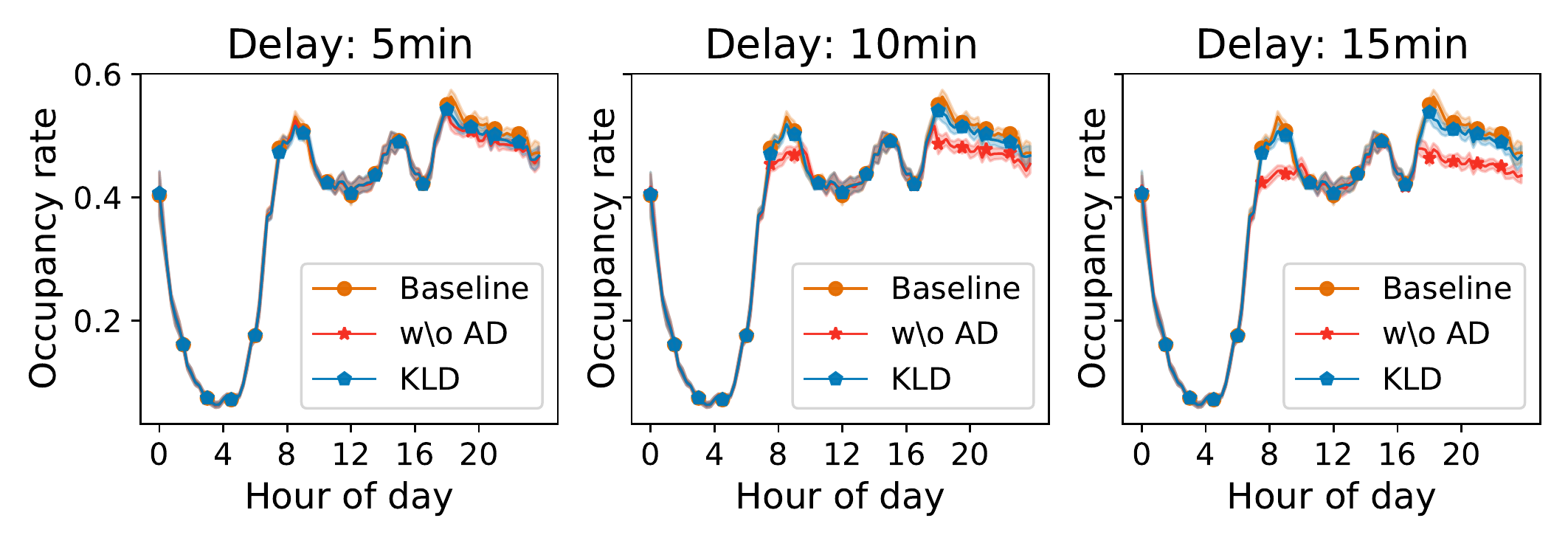}\label{fig:occupancy}}
\caption{Dynamics of electric mobility service during the last (4th) week in simulation: (a) order fulfillment rate and (b) SEV fleet's occupancy rate.}\label{fig:fulfillment and occupancy rate}
\end{figure}

Figs.~\ref{fig:fulfillment}-\ref{fig:occupancy} illustrate the order fulfillment and SEV fleet's occupancy rates under the baseline and the DCA with and without AD. For the scenarios under AD, we only show the results of the KLD, which is observed to be one of the most effective AD techniques in Fig~\ref{fig:tot_revenue}. 
In Fig.~\ref{fig:fulfillment}, distinct differences between the baseline and scenarios without the AD are observed during the morning and evening peaks (e.g., 8 to 10 AM and 6 to 11 PM). In this case, we report a reduction of fulfillment rates of about 20\% during the morning peak and 12\% during the evening peak, yielding fulfillment rates of 80\% and 67\%, respectively. Similarly, the reduction of SEV occupancy rates varies from 2\% to 6\% under the delay of \SI{5}{min} and \SI{15}{min} from 6 PM to 11 PM. 
Under the PCC, we report an improvement in fulfillment rate of up to 8\% and a SEV occupancy rate as high as 5\% under the delay of \SI{15}{min}.

\subsection{Impacts of EVSE}
\begin{figure}[h]
    \centering
    \includegraphics[width=\linewidth]{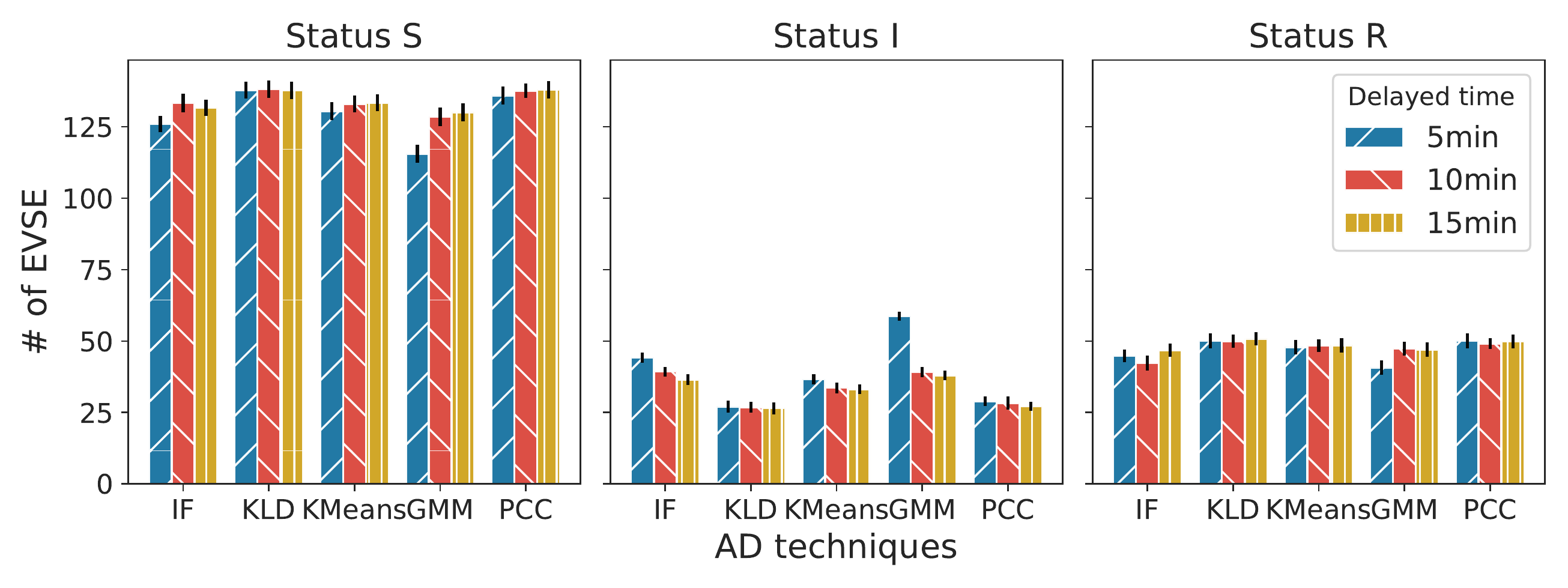}
    \caption{Proportion of EVSE in statuses S, I, and R under five AD techniques.}
    \label{fig:evse_statuses}
\end{figure}

We illustrate in Fig.~\ref{fig:evse_statuses} the SIR proportions of EVSE under different delays and AD techniques during the 4th week.

In general, the numbers of susceptible (status S) EVSE under \SI{5}{min} delay are lower than those under \SI{10}{min} and \SI{15}{min} delay. In particular, under a \SI{5}{min} delay, the IF and GMM are associated with the lowest number of susceptible EVSE, which implies the relatively higher revenue loss in Fig.~\ref{fig:tot_revenue}. At the same time, significantly higher proportions of infectious EVSE and lower proportions of removed EVSE are observed under the IF and GMM, especially with the delay of \SI{5}{min}. This can be explained by the stealthiness of DCA, which can hardly be detected by the AD techniques mentioned above. In addition, the numbers of removed EVSE under \SI{15}{min} delay are reported to be higher than those under \SI{10}{min} delay. In this case, the higher proportion of removed EVSE results in a higher revenue loss, as observed in Fig.~\ref{fig:tot_revenue}, and lower fulfillment rates and occupancy rates in Fig.~\ref{fig:fulfillment and occupancy rate}.

\subsection{Performance of AD Techniques}

\begin{table*}[ht]
    \centering
    \scriptsize
    \caption{Comparison between five AD techniques}
    \label{tab:comparison_detection_techniques}
   \begin{tabular}{c|cccc|cccc|cccc}\hline
   &\multicolumn{4}{c}{\SI{5}{min}} &\multicolumn{4}{c}{\SI{10}{min}} &\multicolumn{4}{c}{\SI{15}{min}} \\\cmidrule(lr){2-5}\cmidrule(lr){6-9}\cmidrule(lr){10-13}
AD     & Accuracy & Precision & Recall & F1 & Accuracy  & Precision  & Recall & F1   & Accuracy & Precision  & Recall & F1  \\\hline
IF     &  0.73(0.01) &  0.82(0.01) &  0.74(0.01) &  0.78(0.01) &   0.80(0.01) &  0.81(0.01) &  0.85(0.01) &  0.83(0.01) &   0.80(0.01) &  0.82(0.01) &  0.86(0.01) &  0.84(0.01) \\\hline
KLD    &  \textbf{0.86(0.05)} &  0.86(0.05) &   \textbf{0.99(0.0)} &  \textbf{0.92(0.03)} &  \textbf{0.87(0.01)} &  \textbf{0.88(0.01)} &   \textbf{0.99(0.0)} &   \textbf{0.93(0.0)} &  0.85(0.01) &  0.86(0.01) & \textbf{0.99(0.00)} &  \textbf{0.92(0.01)} \\\hline
KMeans &  0.78(0.01) &   0.84(0.00) &  0.86(0.01) &   0.85(0.00) &   0.82(0.00) &   0.84(0.00) &   0.91(0.00) &   0.88(0.00) &   0.81(0.00) &   0.84(0.00) &   0.91(0.00) &   0.87(0.00) \\\hline
GMM    &  0.64(0.01) &   0.81(0.00) &  0.57(0.02) &  0.67(0.02) &  0.77(0.01) &  0.79(0.01) &  0.78(0.02) &  0.78(0.01) &   0.79(0.00) &  0.79(0.01) &   0.80(0.01) &  0.79(0.01) \\\hline
PCC    &   0.85(0.00) &   \textbf{0.87(0.00)} &   {0.98(0.00)} &   \textbf{0.92(0.00)} &   {0.85(0.00)} &   {0.86(0.00)} &   {0.98(0.00)} &   {0.92(0.00)} &   \textbf{0.86(0.00)} &   \textbf{0.87(0.00)} &   0.98(0.00) &   \textbf{0.92(0.00)} \\\hline
    \end{tabular}
\end{table*}

\begin{figure*}[ht]
    \centering
    \includegraphics[width=0.9\linewidth]{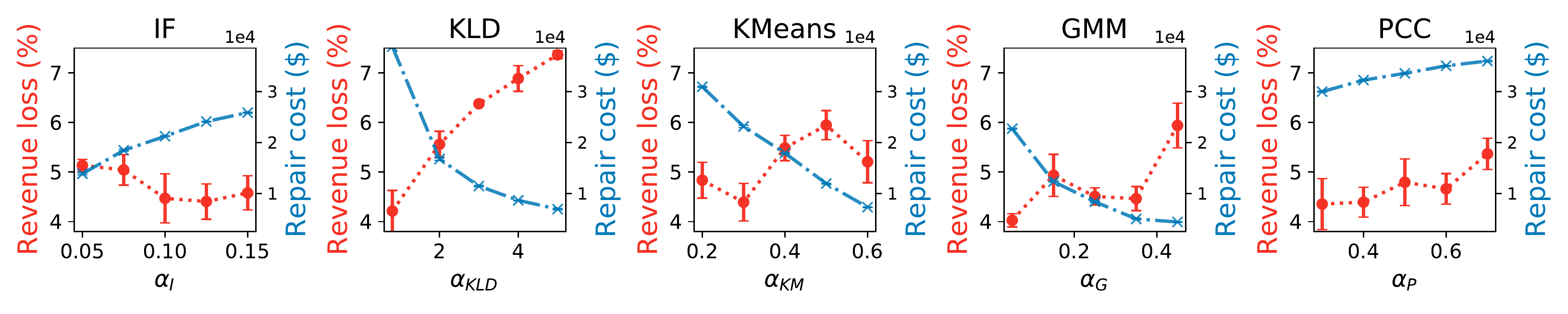}
    \caption{Trade-off between revenue loss and repair cost under a delay of \SI{10}{min} over a three-week period}
    \label{fig:trade_off_revenue_detection}
\end{figure*}

Table~\ref{tab:comparison_detection_techniques} reports the performance between the five AD techniques: accuracy, precision, recall, and F1 score.

We note that the KLD and PCC outperform the other AD techniques under different charging delays, where the accuracy and precision are over 0.86 and F1 scores are at least 0.92. In particular, the recall can reach up to 0.99, which suggests a significantly low miss-detection rate (as low as 1\%). The superior performance of KLD and PCC can be explained by the non-parametric features, where no prior assumptions are made on the sample distribution. In particular, the low-rank approximation under the PCC exhibits the great potential of identifying the anomaly, especially for large-scale problems. In addition to the PCC, we also report good performances under the KMeans. In this case, all metrics are over 0.80 under the \SI{10}{min} and \SI{15}{min} delay. Also, we observe that the precision and recall exceed 0.80 in all cases, indicating a good performance regarding the false-positive and miss-detection rates. 
As a clustering-based technique, high-quality clusters are obtained based on the historical data under normal operation. The outliers are likely to be associated with high anomaly scores (i.e., long distance to the cluster).
\subsection{Sensitivity Analysis}
To better understand the trade-off between repair cost and system revenue loss, we conduct sensitivity analyses on the AD techniques with respect to $\alpha_{(\cdot)}$ (the ranges for the cross-validation procedure were detailed in Table~\ref{tab:cv_for_AD}). We summarize the repair cost and revenue loss under the \SI{10}{min} delay in Fig.~\ref{fig:trade_off_revenue_detection}. 
We first observe that the higher sensitivity levels lead to linearly or even exponentially increasing repair costs, ranging from about \$6,000 (KMeans) to nearly \$40,000 (KLD and PCC). This is because the AD models aim to perform more sensitive DCA detection by grouping more charging events as anomalies, resulting in higher repair costs to send out a technician for repair service. At the same time, the revenue loss (red dotted lines) witnessed an overall reduction with significantly higher repair costs due to more sensitive detection. However, the revenue loss remains at least 4\% in all five AD models regardless of the repair cost, which suggests the robustness of the DCA model.
Observing the trade-off between repair cost and revenue loss, we note that a slight compromise on the revenue loss can result in a significantly lower repair cost, e.g., $\alpha_{I}=0.05,0.075$, $\alpha_{KLD}=1,2$, and $\alpha_{P}=0.3,0.4$. Those trade-off decisions will shed light on the coordinated management of SEV and EVSE for commercial purpose~\cite{blink2022ev}. We highlight that there will always be a proportion of infectious EVSEs serving the SEV fleet, resulting in huge revenue loss regardless of the sensitivity levels of the DCA detection models. Meanwhile, more sensitive detection may not contribute to more successful detection but lead to exponentially increasing repair costs and higher miss-detection rates (proportion of false-positive and false-negative alarms). In this regard, we alert more tailored AD models that can best balance the trade-off between system revenue and repair cost.

\section{Conclusion}\label{sec:conclusion}
This study presents a novel DCA model that can surreptitiously impede the ESMS by delaying the charging service of the SEV fleet. By utilizing the NYC taxi trip data and real-world EVSE locations, we evaluated the system impacts of the DCA on the ESMS in NYC using a self-developed high-fidelity simulation platform. Our results demonstrate a long-term degradation of the ESMS caused by the DCA, with a \SI{10}{min} delay resulting in up to 35\% longer queuing times at EVSE during the daytime (11AM - 6PM) and up to 6.8\% longer average charging times at around 10AM. Furthermore, such a disruption to the charging service leads to an 8\% increase in unfulfilled requests, which results in a 10.7\% (\$311.7) weekly revenue loss per SEV driver. Even with the AD techniques, our results show that the weekly revenue loss remains at a minimum of 3.8\% (\$111.8), along with increased repair costs of up to \$36,000 per week. Therefore, our DCA model highlights a realistic and stealthy cyberattack approach that can chronically harm the ESMS. In conclusion, this study contributes to the field of cybersecurity by introducing a new cyberattack approach and providing insights into the system-level impacts of the DCA on the ESMS.

Future research should focus on incorporating a more realistic EVSE choice model to better understand the impacts of DCA and its cascading failures. For instance, the SEV may detour to a relatively distant EVSE due to the preference~\cite{guo2022modeling}. Additionally, tailored defense strategies should be developed to effectively identify malfunctioned EVSEs, even under minimal disruptions in charging service (e.g., a \SI{2}{min} delay). This is especially relevant to the increasing prevalence of high-wattage EVSEs (e.g., extreme fast chargers~\cite{afdc2022chargingspeed}) and heavy-duty electric trucks in the ESMS. 
Moreover, SEV drivers may have different target SoC levels (e.g., above or below 80\%) due to time constraints, which can result in a non-linear CC-CV charging profile. Future studies should incorporate this factor to simulate more realistic charging behaviors. 
Finally, battery degradation~\cite{zhao2021assessment} is an important consideration that adds another layer of variability to charging duration beyond human-involved activities. Incorporating battery degradation in our future research will make our proposed DCA approach even more robust and applicable to real-world scenarios.

% \section{Author Contributions}
% The authors confirm contribution to the paper as follows: study conception and design: SG, HC, and XQ; data collection: SG and XQ; analysis and interpretation of results: SG and XQ; draft manuscript preparation: SG, HC, MR, and XQ. All authors reviewed the results and approved the final version of the manuscript.

\bibliographystyle{ieeetr}
\bibliography{ref}

% Biography text here without a photo.

% \begin{IEEEbiography}[{\includegraphics[width=1in,height=1.25in,clip,keepaspectratio]{fig1.png}}]{IEEE Publications Technology Team}
% In this paragraph you can place your educational, professional background and research and other interests.\end{IEEEbiography}
\begin{IEEEbiography}[
{\includegraphics[width=1in,height=1.25in,clip,keepaspectratio]{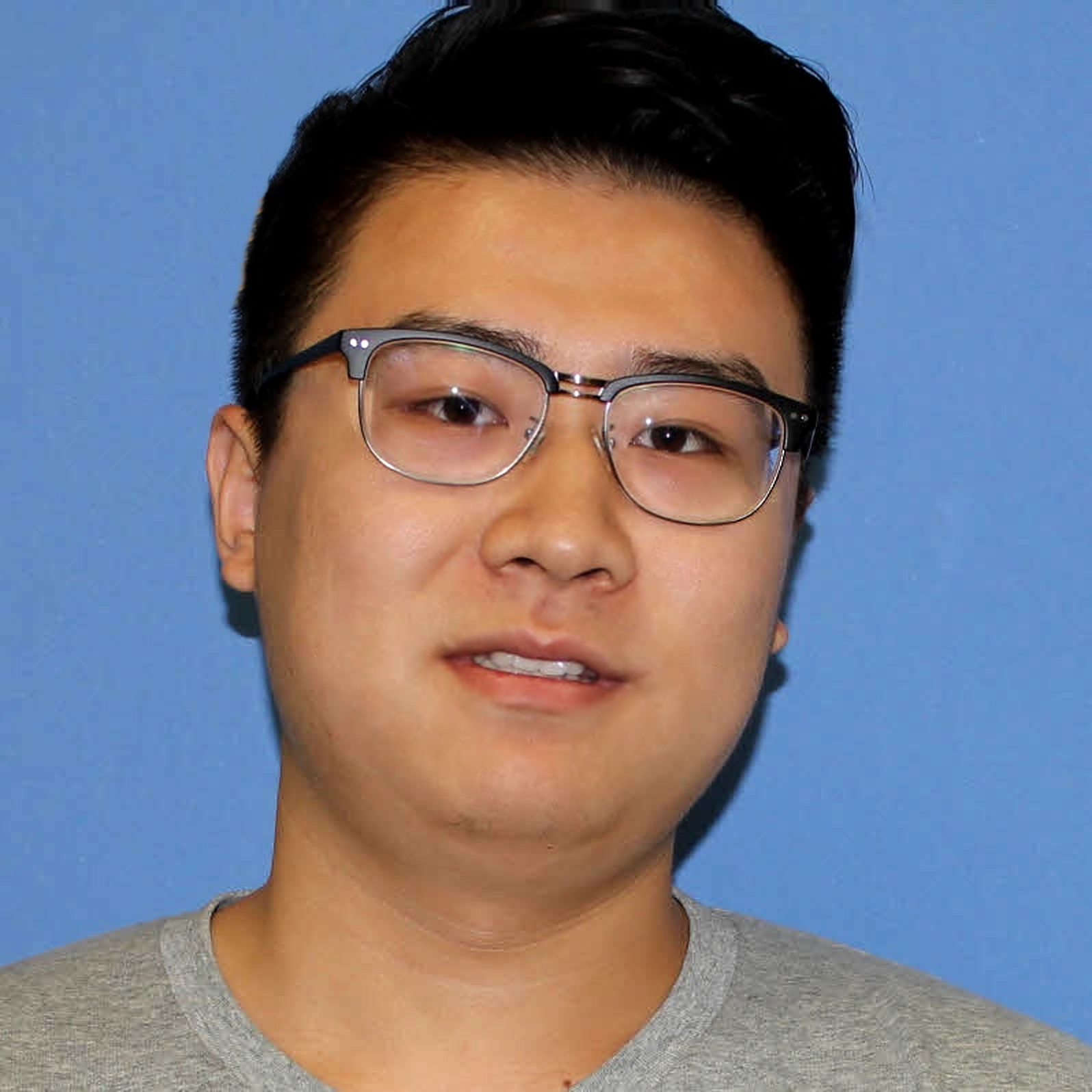}}
]
{Shuocheng Guo} received the B.E degree in civil engineering from Central South University, Changsha, China, and the M.S. degree in transportation engineering from University of Illinois at Urbana and Champaign, Urbana, IL, USA. He is currently working towards the Ph.D. degree in transportation engineering at The University of Alabama, Tuscaloosa, AL, USA. His research interests include optimization, electrified transportation network, and complex network analysis.
\end{IEEEbiography}

\begin{IEEEbiography}
[
{\includegraphics[width=1in,height=1.25in,clip,keepaspectratio]{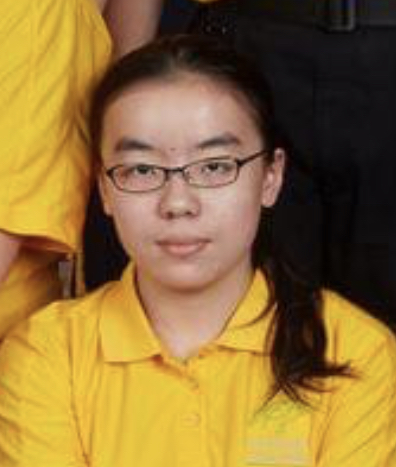}}
]
{Hanlin Chen} received the B.S. degree in Electrical engineering and automation from Huazhong University of Science and Technology, Wuhan, Hubei, China, and the M.S. degree in Mechanical Engineering Technology and Ph.D. degree in Computer and Information Technology from Purdue University, West Lafayette, IN, USA. She is currently a postdoctoral research assistant in Civil Engineering, Purdue University. Her research interests include Cooperative perception, 
Traffic-informed perception, planning on vehicle side, Cybersecurity and Resilience in CDA system and CAV in homeland security.
\end{IEEEbiography}

\begin{IEEEbiography}
[{\includegraphics[width=1in,height=1.25in,clip,keepaspectratio]{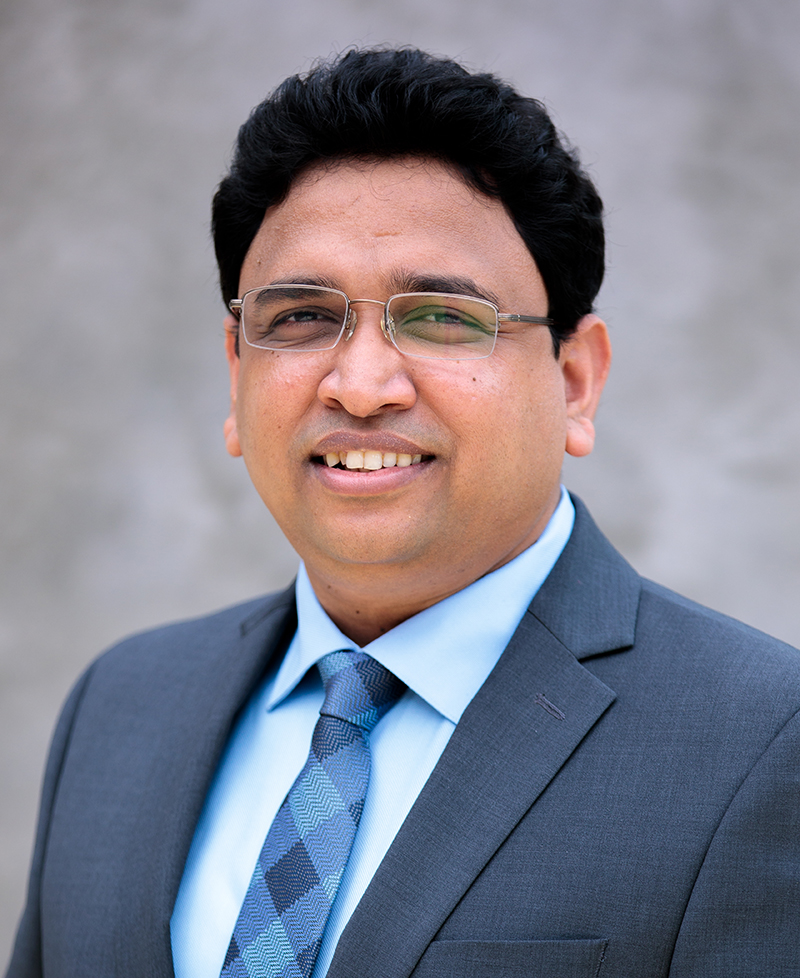}}]
{Mizanur Rahman}  is an assistant professor in the
Department of Civil, Construction and Environmental
Engineering at the University of Alabama,
Tuscaloosa, Alabama. After his graduation in August
2018, he joined as a postdoctoral research fellow for
the Center for Connected Multimodal Mobility
(C$^2$M$^2$), a U.S. Department of Transportation Tier 1
University Transportation Center
(cecas.clenson.edu/c2m2). After that, he has also
served as an Assistant Director of C2M2. His research
focuses on traffic flow theory, and transportation cyber-physical systems for
connected and automated vehicles and smart cities.
\end{IEEEbiography}

\begin{IEEEbiography}
[
{\includegraphics[width=1in,height=1.25in,clip,keepaspectratio]{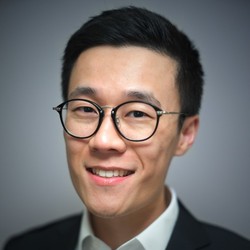}}
]
{Xinwu Qian} received the B.S. degree in transportation engineering from Tongji University, Shanghai,
China, and the M.S. and Ph.D. degrees in transportation engineering from Purdue University, West
Lafayette, IN, USA. He is currently an Assistant
Professor of Civil, Construction, and Environmental
Engineering with The University of Alabama. His
research interests include big data analytics, complex network analysis, and network modeling and optimization.
\end{IEEEbiography}

\end{document}